\documentstyle[aps,epsfig]{revtex}
\begin{document}
\draft
\sloppy
\title{Hartree Fock Calculations in the Density Matrix Expansion Approach}
\author{F. Hofmann and H. Lenske}
\address{Institut f\"ur Theoretische Physik, Universit\"at Gie\ss en\\
	D-35392 Gie\ss en}
\maketitle

%-------------------------------------------------------------------------------------
\begin{abstract}
The density matrix expansion is used to derive a local energy density functional
for finite range interactions with a realistic meson exchange structure. Exchange
contributions are treated in a local momentum approximation. A generalized Slater
approximation is used for the density matrix where an effective local Fermi
momentum is chosen such that the next to leading order off-diagonal term is
canceled. Hartree--Fock equations are derived incorporating the momentum
structure of the underlying finite range interaction. For applications a density
dependent effective interaction is determined from a G-matrix which is
renormalized such that the saturation properties of symmetric nuclear matter are
reproduced. Intending applications to systems far off stability special attention
is paid to the low density regime and asymmetric nuclear matter. Results are
compared to predictions obtained from Skyrme interactions. The ground state
properties of stable nuclei are well reproduced without further adjustments of
parameters. The potential of the approach is further exemplified in calculations
for A=100$\dots$140 tin isotopes. Rather extended neutron skins are found beyond
$^{130}$Sn corresponding to solid layers of neutron matter surrounding a core of
normal composition.
\end{abstract}
%-------------------------------------------------------------------------------------

\pacs{21.60-n,21.60Jz,21.10.Dr}

%-------------------------------------------------------------------------------------
\section{Introduction}

Studies of nuclei far off stability have given a new impact to nuclear structure
physics. Exotic nuclei allow for the first time to test nuclear theory at extreme
isospin and, close to the driplines, also at low density. In the past nuclear
models have been derived almost exclusively from well bound and charge symmetric
stable nuclei. Hence, structure investigations far off stability are providing
important extensions into hitherto unexplored regions of the nuclear landscape.
It is apparent that such studies involve far reaching extrapolations of nuclear
models much beyond the range where they were initially developed. The weaker
binding of strongly asymmetric systems enhances interactions between bound and
unbound configurations and the regime of open quantum systems might be
approached. In extreme dripline nuclei as e.g. $^8$B \cite{b8} and $^{11}$Li
\cite{li} evidence for a strong competition of interactions between individual
nucleons and mean--field dynamics is found.

At present nuclear structure theory is making first exploratory steps into the
new regions far off $\beta$-stability. Experience shows that seemingly well
established models describing  stable nuclei equally well are leading to strong
deviations when applied to exotic systems. A particular striking feature are the
variations in predictions of binding energies and other ground state properties
at large charge asymmetry. The results indicate that the isovector properties of
interactions are not well enough understood in order to reproduce or predict the
location of driplines. A widely applied method in non--relativistic structure
calculations is the Skyrme approach \cite{skyrme}. Although very successful in
stable nuclei, it is an open question to what extent the restricted set of
operators used in the Skyrme energy functional is sufficiently accounting for the
conditions in the new mass regions. The transparent structure and the good
applicability is a clear advantage of the Skyrme model. It is also flexible
enough for adjusting the already incorporated parameters or introducing
additional ones in order to describe new data. However, a purely empirical
approach might be hampered by increasing uncertainties when approaching the
limits of stability.

The intention of the present work is to indicate an alternative approach which is
closer to a more microscopic description but still retains to a large extent the
advantages of using a local energy functional as in Skyrme theory. A more
fundamental description requires to use microscopically derived in--medium
interactions as e.g. the Brueckner G--matrix accounting for medium effects on the
level of ladder diagrams \cite{gmat,Koehler}. Theoretically, such a full scale
many-body calculation is not feasible as a standard approach to finite nuclei,
except for selected cases \cite{mueth1,kuemmel}. A successful program is to use
instead a G--matrix from infinite nuclear matter and to apply it in the local
density approximation (LDA) to finite nuclei \cite{gmat}. The LDA has been used
succesfully in non--relativistic \cite{negvaua,negrmp,spin} and also covariant
fieldtheoretical models \cite{Anast,HS,tHM,MMB,BM,BT,LF1,LF2}. However, the
finite range character of a realistic G--matrix is well known to introduce
complications on the numerical level by leading to a system of
integro--differential equations \cite{gogny}. A way around this problem was
discussed some time ago by Negele \cite{negvaua} but rarely applied later. The
numerical complications arise because of the non--locality of exchange
contributions to the HF--field. For a finite range interaction they involve the
full one--body density matrix rather than the local densities needed for the
direct parts and contact interactions of Skyrme type. Negele introduced an
operator-valued expansion of the non--local parts of the density matrix. The
contraction with the HF--ground state configuration leads to a systematic
expansion in terms of the local one-body density. The non--localities are
contained in functions where the lowest order contribution coincides with the
well known Slater approximation \cite{negvaua,baduri}. This density matrix
expansion (DME) allows to integrate out the non--localities of the exchange
terms.

The approach presented below follows in spirit closely the original work of
Ref. \cite{negvaua,negrmp}. In section \ref{2} the basic steps of the DME are
summarized. Different to Negele we choose an effective local ''Fermi'' momentum
q$^2_F$({\bf r}) such that the next to leading order term of the DME series is
canceled exactly. q$^2_F$  is determined by the kinetic energy density and, in
addition, the Laplacian of the one--body density. For an infinite system the
conventional expression for the Fermi momentum is recovered. The DME results are
then used to define a local energy density functional. Since it originates from
an effective interaction, assumed to be given independently e.g. by Brueckner
calculation, the parameters of the model are fixed. By variation of the energy
functional HF--equations are derived. With our choice for q$_F$ the kinetic
energy part of the single particle Schroedinger equation receives additional
contributions. Thus, we obtain consistently effective masses m$^*_{p,n}$(r) for
protons and neutrons, respectively, which are completely determined by the
momentum and isospin structure of the original interaction. An important
extension over the Skyrme approach is that the self--consistent local momentum
q$_F$ contributes in arbitrarily high order to the exchange terms which is
especially important at high and low densities.

Applications to infinite symmetric and asymmetric nuclear matter are discussed in
section \ref{3}. We start by choosing a somewhat simplified effective
interaction. The three Yukawa parameterization of a G--matrix calculated
initially for $^{16}$O by Toki et al.\cite{toki,bertsch} is used. This so--called
M3Y interaction has been widely and successfully applied in nuclear reaction
\cite{Satch,oertzen} and structure calculations \cite{spin,Eck,somm}. The former
studies, especially for nuclear reactions, indicate that this interaction has a
realistic momentum structure including a long range one pion--exchange tail. It
is especially trustable at low nuclear densities around $1/3$ of the saturation
density. A density dependence is lacking and it is known that the original M3Y
parameterization does not lead to saturation of nuclear matter. As other authors
before \cite{oertzen}, we restore the density dependence by fitting the
properties of symmetric nuclear matter at saturation including the binding
energy, equilibrium density, compressibility and, in addition, also the symmetry
energy. This allows to fix the density dependence of the isoscalar and isovector
vertices in the spin--scalar particle--hole interaction channels. We thus have
determined a renormalized density dependent three Yukawa (D3Y) interaction which
we expect to resemble closely the momentum structure of a G--matrix but with an
effective strength fitted to nuclear matter properties \cite{bethe}. The
in--medium particle--particle interaction is obtained by the appropriate
transformation and used in pairing calculations. At low densities the strength of
the free singlet--even NN T--matrix is approached. Results for symmetric and
asymmetric nuclear matter are compared to calculations with empirical Skyrme
interactions. Close to saturation a good agreement is found but deviations occur
away from that point.

In section \ref{4} the model is applied to finite nuclei. The parameters are kept
fixed as obtained from the nuclear matter fits. Binding energies, charge
densities and radii of stable nuclei ($^{16}$O,$^{40,48}$Ca,$^{90}$Zr and
$^{208}$Pb) are well described. The good agreement confirms the approach and
makes applications to unstable nuclei meaningful. HF calculations for Sn isotopes
lead to a good description of the binding energies for A$\geq$120 but a slightly
too weak binding is found in the lighter nuclides. The calculations predict
rather thick neutron skins for A$\ge$130. The paper closes in section \ref{5}
with a summary and conclusions.

%-------------------------------------------------------------------------------------
\section{Density--Matrix Expansion and Hartree--Fock Theory} \label{2}

The total Hartree--Fock energy of a nucleus with A nucleons and the single
particle states $|k\rangle$ is given by the kinetic energy
\begin{equation}
	T = \sum_{k \leq A} \langle k | \frac{p^{2}}{2 m} | k \rangle = \int
	d^{3}r \,\frac{\hbar^{2}}{2 m}\,\tau({\bf r}) = \int d^{3}r
	\,\frac{\hbar^{2}}{2 m}\, \left\{ \tau_{p}({\bf r}) + \tau_{n}({\bf r})
	\right\},
\end{equation}
where $q=n,p$ denotes protons and neutrons and $\tau_{q}({\bf r})$ the kinetic
energy densities. The potential energy of a two-body interaction $V$
\begin{equation}
	\langle \bar V \rangle = \left[ \sum_{k_{1} k_{2}} \langle k_{1} k_{2}| V
	| k_{1} k_{2} \rangle - \langle k_{1} k_{2}| V | k_{2} k_{1} \rangle
	\right]
\end{equation}
is expressed in terms of interactions for like ($q=q'$) and unlike ($q\neq q'$)
particles as defined in App. \ref{app}
\begin{eqnarray} \label{totpot}
	\langle \bar V \rangle	& = & \langle V^{d} \rangle + \langle V^{e}
	\rangle = \sum_{q q'=p,n} \left\{\langle V_{qq'}^{d} \rangle  + \langle
	V_{qq'}^{e} \rangle \right\} \\ \nonumber & = & \sum_{q q'} \int
	d^{3}r_{1} \int d^{3}r_{2} \, \left\{ \rho_{q}({\bf r}_{1})
	\rho_{q'}({\bf r}_{2}) V_{qq'}^{d}({\bf r}_{12}) + \rho_{q}({\bf
	r}_{1},{\bf r}_{2}) \rho_{q'}({\bf r}_{1},{\bf r}_{2}) V_{qq'}^{e}({\bf
	r}_{12}) \right\}
\end{eqnarray}
where direct and exchange contributions are indicated separately by the index $d$
resp. $e$. Here, ${\bf r}_{12} = {\bf r}_{1} - {\bf r}_{2}$ is the relative
coordinate and the density matrices are defined as
\begin{eqnarray}
	\rho_{q}({\bf r}_{1}, {\bf r}_{2}) & = & \sum_{k \ \sigma}
	\phi_{k}^{*}({\bf r_{1}}, \sigma, q) \phi_{k} ({\bf r}_{2}, \sigma, q) \\
	\nonumber \rho_{q}({\bf r}) & = & \rho_{q}({\bf r}, {\bf r})
\end{eqnarray}
where $\phi_{k} ({\bf r}_{2}, \sigma, q)$ denotes single particle wave functions.

%-------------------------------------------------------------------------------------
\subsection{Density Matrix Expansion}

The exact treatment of the exchange term in the Hartree-Fock functional would,
after variation, lead to a coupled system of integro-differential equations. A
numerically simplified and theoretically more transparent approach is obtained
using the density-matrix expansion (DME) which was originally invented by Negele
and Vautherin \cite{negvaua}. In the DME approach one first transforms to
center-of-mass and relative coordinates by introducing ${\bf r} =
\frac{{\bf r}_{1} + {\bf r}_{2}}{2}$ and ${\bf s} = {\bf r}_{2} - {\bf r}_{1}$.
Then the density matrix is formally expanded into a Taylor series in ${\bf s}$. A
particular simple result is obtained if the time-reversed orbitals in the nucleus
are filled pairwise \cite{negvaua}. Therefore, the DME approach is most suitable
for even-even nuclei. As shown in \cite{negvaua}, averaging over spin directions
and resummation leads to a series expansion where the first two terms are
\begin{eqnarray} \label{dme2}
	\lefteqn{\rho \left({\bf r} + \frac{{\bf s}}{2},{\bf r} - \frac{{\bf s}}
	{2}\right) = \rho_{SL}(s q_{F})\,\rho({\bf r})} \\ & & \hspace{2cm} + \,
	\, \frac{35}{2s q_{F}^{3}}j_{3}(s q_{F})\, \left[ \frac{1}{4} {\bf
	\nabla}^{2} \rho({\bf r}) - \tau({\bf r}) +
	\frac{3}{5}q_{F}^{2}\,\rho({\bf r}) \right] + \cdots \nonumber
\end{eqnarray}
and in the leading order term the well known Slater approximation
\cite{negvaua,baduri} is recovered. Here, the Slater density $\rho_{SL}(s q_{F})
= \frac{3}{s q_{F}} j_{1}(s q_{F})$ is given by the spherical Bessel function
$j_{n}(x)$ of the order $n=1$.

The quality of the expansion depends on the choice of q$_{F}^{2}({\bf r})$ which
should obviously correspond to the average relative momentum between the two
interacting particles. Different to Ref. \cite{negvaua} we choose
\begin{equation} \label{modfermi}
	q_{F}^{2}({\bf r}) = \frac{5}{3} \frac{\tau({\bf r}) - \frac{1}{4} {\bf
	\nabla}^{2} \rho({\bf r})}{\rho({\bf r})}.
\end{equation}
which is seen to cancel exactly the next to leading order term in Eq.(\ref{dme2})
and has the property $\int d^{3}r \, \rho({\bf r}) q_{F}^{2}({\bf r}) = 5/3 \int
d^{3}r \, \tau({\bf r})$.

With this choice the DME reduces to the well known Slater approximation
\label{dichte} $\rho \left({\bf r} + \frac{{\bf s}}{2},{\bf r} - \frac{{\bf
s}}{2}\right) = \rho_{SL}(sq_{F})\,\rho({\bf r})$ of the exchange term but now
with a modified Fermi momentum that accounts for surface corrections to the Fermi
momentum of the Thomas Fermi model up to the second order. The advantage of this
method is that the description of the exchange potential is simplified and can be
treated in a way similar to the nuclear matter calculation. It is easily seen
that our choice of the Fermi momentum reduces in infinite nuclear matter to the
conventional expression $k_{F}=(3/2 \pi^{2}\rho)^{1/3}$.

Once the relative and the center-of-mass coordinates are separated and the
integration over the relative coordinate ${\bf s}$ can be immediately performed
analytically. The potential energy of the exchange part finally simplifies to:
\begin{eqnarray}
	\langle V^{e} \rangle & = &  \sum_{qq'} \int d^{3}r \, \rho_{q}({\bf r})
	\rho_{q'}({\bf r}) \tilde V_{qq'}^{e}({\bf r})
\end{eqnarray}
In this equation the strength of the interactions are defined as
\begin{equation} \label{delta}
	\tilde V_{qq'}^{e}(q_{F_{q}}, q_{F_{q'}}; {\bf r}) = \int d^{3}s
	V_{qq'}^{e}(s) \rho_{SL}(s q_{F_{q}}({\bf r})) \rho_{SL}(s q_{F_{q'}}
	({\bf r}))
\end{equation}
In this way the exchange term of a finite range interaction is mapped to an
effective contact interaction with a local strength depending on the effective
local Fermi momentum, Eq.(\ref{modfermi}), at the place of the interaction
\begin{equation}
	V^{e}_{qq'}(s, {\bf r}) \rightarrow \delta^{3}({\bf s}) \, \tilde
	V_{qq'}^{e}(q_{F_{q}}({\bf r}), q_{F_{q'}}({\bf r})).
\end{equation}
We are now at the position to reformulate the theory in terms of a local energy
density functional
\begin{equation} \label{eint}
	H({\bf r}) = H_{T}({\bf r}) + H_{V^{d}}({\bf r}) + H_{V^{e}}({\bf r})
\end{equation}
with the total energy
\begin{equation}
	E = \int d^{3}r \, H({\bf r}) = T + \frac{1}{2} \langle \bar V \rangle .
\end{equation}
Comparing this to the equations derived before one finds:
\begin{eqnarray}
	H_{T}({\bf r}) & = & \frac{\hbar^{2}}{2 m}\,\left[ \tau_{p}({\bf r}) +
	\tau_{n}({\bf r}) \right] \\ H_{V^{d}}({\bf r}) & = & \frac{1}{2} \,
	\sum_{q q'} \rho_{q}({\bf r}) U_{qq'}^{d}({\bf r}) \\ H_{V^{e}}({\bf r})
	& = & \frac{1}{2} \, \sum_{q q'} \rho_{q}({\bf r}) U_{qq'}^{e}({\bf r})
\end{eqnarray}
Single particle potentials $U({\bf r}, \rho)$ have been introduced as e.g.
\begin{eqnarray} \label{sppd}
	U_{qq'}^{d}({\bf r}) & = & \int d^{3}r' \ \rho_{q'}({\bf r'})
	\,V_{qq'}^{d}({\bf r},{\bf r'}) \\ U_{qq'}^{e}({\bf r}) & = &
	\rho_{q'}({\bf r}) \tilde V_{qq'}^{e}(q^{2}_{F_{q}}, q^{2}_{F_{q'}}; {\bf
	r}) \label{sppe}
\end{eqnarray}

%-------------------------------------------------------------------------------------
\subsection{The Hartree--Fock Equations} \label{hfe}
In our case, the total energy depends in a complex, nonlinear way on the nuclear
density $\rho$, the kinetic energy density $\tau$ and ${\bf
\nabla}^{2} \rho$. By using the DME we are able to simplify the exchange term.
Our choice of the Fermi momentum $q_{F}(\rho, \tau, {\bf \nabla}^{2} \rho)$,
Eq.(\ref{modfermi}), leads to an additional functional dependence on the kinetic
energy and the Laplacian of the density and therefore takes into account
dynamical and surface corrections. For the variation of the HF energy functional,
Eq.(\ref{eint}), we express $\rho$, $\tau$ and $q_{F}^{2}$ by $\phi_{k}$ and its
derivatives:
\begin{eqnarray} \label{ac1}
	\rho({\bf r}) & = & \sum_{k} \left| \phi_{k}({\bf r}) \right|^{2} \\
	\tau({\bf r}) & = & \sum_{k} \left| {\bf \nabla} \phi_{k}({\bf r})
	\right|^{2} \\ q_{F}^{2}(\phi_{k}, {\bf \nabla} \phi_{k}, {\bf
	\nabla}^{2} \phi_{k}) & = & \frac{5}{6} \frac{1}{\rho} \sum_{k} \left\{
	\left| {\bf \nabla} \phi_{k} \right|^{2} - \frac{1}{2} \left(
	\phi_{k}^{*} {\bf \nabla}^{2} \phi_{k} + \phi_{k} {\bf \nabla}^{2}
	\phi_{k}^{*} \right) \right\} \label{ac4}
\end{eqnarray}
The HF variational equation
\begin{equation}
	\int d^{3}r\, \delta \left( H({\bf r}) - \sum_{k} \left| \phi_{k}({\bf
	r}) \right|^{2} \epsilon_{k} \right) = 0
\end{equation}
leads after partial integration	 to
\begin{equation} \label{eqc}
	\delta E = \sum_{k} \int d^{3}r \,\delta \phi_{k} \left( \frac{\partial
	H}{\partial \phi_{k}^{*}} - {\bf \nabla} \frac{\partial H}{\partial {\bf
	\nabla} \phi_{k}^{*}} - {\bf \nabla}^{2} \frac{\partial H}{\partial {\bf
	\nabla}^{2} \phi_{k}^{*}} - \phi_{k} \epsilon_{k} \right)
\end{equation}

For the further calculations we assume that the nucleon--nucleon interaction has
an intrinsic density dependence resulting from the medium dependence of the
meson--nucleon vertex that can be parameterized by vertex corrections
$g_{\gamma}(\rho)$. A model will be specified in the next section.

Then, the partial derivative in Eq.(\ref{eqc}) with respect to $\phi_{k}$ is
written as
\begin{equation} \label{v0}
	\frac{\partial H} {\partial \phi_{k}^{*}}  = \frac{\partial H} {\partial
	\rho} \frac{\partial {\rho} } {\partial \phi_{k}^{*}} + \frac{\partial H}
	{\partial g_{\gamma}(\rho)} \frac{\partial {g_{\gamma}(\rho)} }
	{\partial{\rho}} \frac{\partial {\rho} } {\partial \phi_{k}^{*}} +
	\frac{\partial H} {\partial q_{F}^{2}} \frac{\partial q_{F}^{2}}
	{\partial \phi_{k}^{*}}
\end{equation}
The first term is the ordinary HF potential, the other terms are rearrangement
terms resulting from the intrinsic density dependence of the interaction.
Analogously the other partial derivatives are expressed as
\begin{eqnarray} \label{v1}
	\frac{\partial H}{\partial {\bf \nabla} \phi_{k}^{*}}  & = &
	\frac{\partial H}{\partial \tau} \frac{\partial \tau} {\partial {\bf
	\nabla} \phi_{k}^{*}}  + \frac{\partial H}{\partial q_{F}^{2}}
	\frac{\partial q_{F}^{2}} {\partial {\bf \nabla} \phi_{k}^{*}} \\
	\label{v2} \frac{\partial H}{\partial {\bf \nabla}^{2} \phi_{k}^{*}} & =
	& \frac{\partial H}{\partial q_{F}^{2}} \frac{\partial q_{F}^{2}}
	{\partial {\bf \nabla}^{2} \phi_{k}^{*}}
\end{eqnarray}
Here, the second term in Eq.(\ref{v1}) gives rise to an effective mass. Using the
above definitions and plugging in everything into Eq.(\ref{eqc}) leads to the
Schroedinger-equation
\begin{equation} \label{main}
	\left\{ - {\bf \nabla} \frac{\hbar^{2}}{2m_{q}^{*}({\bf r})} {\bf \nabla}
		+ U^{HF}_{q}({\bf r}) + U^{R}_{q}({\bf r}) \right\}
		\phi_{k_{q}}({\bf r}) = \epsilon_{k_{q}} \phi_{k_{q}}({\bf r}).
\end{equation}
The Hartree-Fock potential is given by
\begin{equation}
	U^{HF}_{q}({\bf r}) = \sum_{q'} \left[ U^{d}_{qq'}({\bf r}) +
	U^{e}_{qq'}({\bf r}) \right]
\end{equation}
and the effective masses are
\begin{equation} \label{emass}
	m^{*}_{q}({\bf r}) = \frac{m}{1 + \frac{2m}{\hbar^{2}} \frac{5}{6} \left[
	\frac{\partial U^{e}_{qq}}{\partial q_{F_{q}}^{2}} + 2	\frac{\partial
	U^{e}_{qq'}}{\partial q_{F_{q}}^{2}} \right] } \quad .
\end{equation}
The effective mass results from the dependence of the Fermi momentum on the
kinetic energy density $\tau$ in the exchange term. It is a direct consequence of
our choice of the Fermi momentum and the finite range nature of the interaction.

The variation of the density dependent vertex functions $g_{\gamma}(\rho)$ and of
the Fermi momentum in the Fock energy density leads to an additional
rearrangement potential
\begin{eqnarray} \label{rpot}
	U^{R}_{q}({\bf r}) & = & \frac{\partial \left( H_{V^{d}} +
	H_{V^{e}}\right)}{\partial g_{\gamma}(\rho)} \frac{\partial
	g_{\gamma}(\rho)}{\partial \rho} \\ \nonumber & & - \frac{1}{2}\,
	q_{F_{q}}^{2} \left[ \frac{\partial U^{e}_{qq}}{\partial q_{F_{q}}^{2}}
		+ 2 \frac{\partial U^{e}_{qq'}}{\partial q_{F_{q}}^{2}} \right] -
	\frac{5}{24} \nabla^{2} \left[ \frac{\partial U^{e}_{qq}}{\partial
	q_{F_{q}}^{2}}
		+ 2 \frac{\partial U^{e}_{qq'}}{\partial q_{F_{q}}^{2}} \right]
\end{eqnarray}
which is of importance for the mean--field or single particle potential in the HF
equations. It results from the density dependence of the NN interaction and is
known to describe (static) polarization of the background medium
\cite{LF2,koehler,revmod}.

%-------------------------------------------------------------------------------------
\section{Nuclear Matter} \label{3}

%-------------------------------------------------------------------------------------
\subsection{The Nucleon--Nucleon Interaction} \label{b}

A well established approach to derive in--medium interactions from the
NN--interaction in free space is Brueckner theory. However, the ladder
approximation, even if refined by going beyond the quasiparticle
approximation as e.g. in Ref.\cite{dejong}, is generally found to miss
the empirical saturation properties of nuclear matter at least when
including two--body correlations only. For the purpose of the present work
we therefore use a semi-phenomenological approach.

We start from the M3Y G--matrix \cite{toki,bertsch} obtained originally for
$^{16}$O. The M3Y G-matrix is parameterized by a superposition of three Yukawa
functions where the ranges were chosen to represent the long-range tail of the
one-pion-exchange potential (OPEP) and medium and short-range parts. The shorter
ranges correspond to boson masses $m=490$ MeV and $m=780$ MeV and thus mimic
$\sigma$, $\omega$ and $\rho$ meson exchange interactions. Here, we use the
parameter set obtained from the Paris NN-potential \cite{toki} which is given in
Tab. \ref{tm3y}.

Because of the lack of a density dependence the M3Y interaction can be
considered as realistic only for a small range of nuclear densities
\cite{satchlove} being centered around one third of the saturation
density. In connection with optical potential and nuclear reaction
studies various authors, e.g. Kobos et al. \cite{kobos} and more
recently Khoa and von Oertzen \cite{oertzen}, have noted that the "bare"
M3Y interaction leads  at high densities to the collapse of nuclear
matter but by supplementing a phenomenological density dependence the
correct binding properties could be restored. In spirit we apply the
same approach by introducing density dependent vertices fitted to the
saturation properties of infinite symmetric nuclear matter.

In the former works only the medium dependence of the isoscalar--spin scalar
channel was considered \cite{oertzen,kobos}. Here, we choose a more general
description and include a density dependence also in the isovector channel.  We
introduce density dependent isoscalar and isovector vertex functions
f$_{0,\tau}(\rho)$, respectively, multiplying the corresponding ''bare'' M3Y
vertices as a whole but leaving the intrinsic momentum structure untouched.
Hence, we assume that the many--body and the momentum structures are separable in
which we follow the observations made in determining effective in--medium
meson--nucleon coupling constants from Brueckner calculations
\cite{MMB,BM,BT,HW}.

Only central interactions are considered. Aiming at applications to asymmetric
systems it is suitable to rewrite the central singlet and triplet even and odd
components  ($V_{SE},V_{TE},V_{SO},V_{TO}$) in terms of interactions for like
(proton-proton, neutron--neutron) and unlike (neutron--proton) particles. We then
consider the interaction in the "particle--hole" channel which eventually
determines the Hartree--Fock mean--field. Multiplying the spin and isospin
exchange operators onto the interaction effective direct (d) and exchange (e)
interactions are obtained \cite{Satchex}.With the relations given in App.
\ref{app} the isoscalar and isovector interactions are then obtained as
\begin{eqnarray}
    V_{0}^{d/e} & = & \frac{1}{4} (V_{pp}^{d/e} + V_{pn}^{d/e} + V_{np}^{d/e}
    + V_{nn}^{d/e}) \\ V_{\tau}^{d/e} & = & \frac{1}{4} (V_{pp}^{d/e} -
    V_{pn}^{d/e} - V_{np}^{d/e} + V_{nn}^{d/e})
\end{eqnarray}
and medium dependent vertex renormalizations f$_{0,\tau}(\rho)$ are
introduced
\begin{equation} \label{dens} \label{vert}
V_{0,\tau}^{*d/e}(\rho)=f_{0,\tau}(\rho) V_{0,\tau}^{d/e}
\end{equation}
taken to be the same in the direct and the exchange channel. Back transformation
to the singlet/triplet representation (see App. \ref{app}) is used to define the
corresponding in--medium particle--particle interaction. By this procedure we
retain the basic relations of particle-hole and particle-particle interactions.

The question arises how to choose the functional form of the vertices. As a
guidance we refer to Brueckner theory and express the G--matrix R in terms of the
free NN scattering matrix K and subtract the Pauli--blocked contributions from
inside the Fermi sphere \cite{LF2,FeWa}:
\begin{equation}\label{gmat}
R(k,k')=K(k,k')-P\int d^3qK(k,q)G_F(q)R(q,k')
\end{equation}
where G$_F$ denotes the two-body propagator inside the Fermi sphere. Dependencies
on the two particle center--of--mass momentum and energy have been omitted. For
the coupled partial wave channels Eq.(\ref{gmat}) leads to a linear system of
equations. Without attempting to solve Eq.(\ref{gmat}) exactly an estimate of the
density dependence is obtained by choosing k=k$_F$ and, in addition, also q=k$_F$
in the interactions beneath the integral. This amounts to the reasonable
assumption that the momentum dependence of K and R over the Fermi sphere is much
weaker than the variation of the (singular) quantity G$_F$. The principal value
integral over G$_F$ can be performed analytically and one finds that the medium
corrections are given in leading order by the density of states at the Fermi
surface, $<G_F>\sim$N$_F$=mk$_F/(\pi\hbar)^2$. ($<G_F>$ depends in fact also on
the two--particle center--of--mass momentum and energy which we assume to be
fixed to appropriate average values). Thus, as an approximation to
Eq.(\ref{gmat}) we find for the half--off shell G--matrix
\begin{equation}\label{gmata}
R(k_F,k)\simeq\frac{1}{1+K(k_F,k_F)<G_F>}K(k_F,k) \quad .
\end{equation}

Next, we consider Eq.(\ref{gmat}) for the case that a solution R$_0$ at a Fermi
momentum k$^0_{F}$ is known. In Eq.(\ref{gmata}) this corresponds to insert in
the denominator the subtracted Green function $<G_F>-<G^0_{F}>$. Since G$^0_{F}$
is a constant and K is weakly varying we may rewrite Eq.(\ref{gmata}) as
\begin{equation}\label{gmatb}
R(k_F,k)\simeq\frac{1}{1+s(k^0_{F})R_0(k_F,k_F)<G_F>}s(k^0_{F})R_0(k_F,k)
\end{equation}
and a scaling factor
\begin{equation}\label{sfac}
s(k^0_{F})=\frac{1}{1-R_0(k^0_{F},k^0_{F})<G^0_{F}>}
\end{equation}
is found.

These considerations suggest to use for the vertex functions ($\gamma=0,\tau$)
the ansatz
\begin{eqnarray} \label{vertex_fit}
f_{\gamma}(\rho) =  s_{\gamma}\left(1+\sum^{N_\gamma}_{n=1}{ a^\gamma_n
z^{n\beta}}\right)
\end{eqnarray}
where z=$\rho/\rho_0$ and the base exponent is fixed to $\beta$=1/3. The
coefficients a$^\gamma_n$ are to be determined by the fit to nuclear matter
properties. As seen below nuclear matter allows a first (N$_\tau$=1) and third
(N$_0$=3) order approximation for f$_\tau$ and f$_0$, respectively. The overall
scaling factors s$_{\gamma}$ renormalize the interaction strength asymptotically
for $\rho \rightarrow 0$ such that with an appropriate readjustment of the a$_n$
coefficients the nuclear matter saturation properties are unaffected. From the
above discussion it is seen that the scaling factors relate the interaction
vertices to the initial renormalization point k$_{F}^{0}$. In the following,
s$_{\gamma}$ and a$^\gamma_n$ are treated as empirical parameters.

The ground state energy of nuclear matter is
\begin{eqnarray} \label{ehfenergy}
    E & = & \sum_{ \alpha} \sum_{k \le k_{F}} \, \frac{\hbar^{2} k^{2}}{2m} +
    \frac{1}{2} \, \sum_{ \alpha \alpha'}  \sum_{k \le k_{{F}}} \sum_{k' \le
    k'_{{F}}} \, \langle \alpha {\bf k}, \alpha' {\bf k}' | \bar{\,V \,}|
    \alpha {\bf k}, \alpha' {\bf k}' \rangle
\end{eqnarray}
where $\bar{V}$ includes antisymmetrization as discussed above and $\alpha$
represents the spin an isospin quantum numbers $\sigma$ and $q$. In nuclear
matter the single particle wave functions $|\alpha{\bf k} \rangle$ are plane
waves and the Fermi momenta $k_{F_{q}}$ of protons and neutrons, respectively, is
related to the corresponding densities $\rho_{q}$ by
$k_{F_{q}}=(3\pi^{2}\rho_{q})^{1/3}$. The total binding energy per particle in
nuclear matter is obtained as:
\begin{eqnarray}\label{easym}
    \frac{E(\rho, \rho_{\tau})}{A} & = & \frac{3}{5} \frac{ \tau_{p} \rho_{p}
    + \tau_{n}	\rho_{n} }{\rho} \\ \nonumber & + & \frac{1}{2 \rho} \,
    \sum_{q q'} \rho_{q} \rho_{q'} \left[ \tilde V_{qq'}^{d}(\rho) + \tilde
    V_{qq'}^{e}(\rho; k_{F_{q}}, k_{F_{q'}}) \right].
\end{eqnarray}
Here, $\rho=\rho_{n}+\rho_{p}$ and $\rho_{\tau}=\rho_{n}-\rho_{p}$ are the
isoscalar and isovector density, respectively, and $\tau_{q} =
\frac{\hbar^{2} k_{F_{q}}^{2}}{2m}$ denotes the Fermi energy of the
protons and neutrons. The effective density dependent interactions appearing in
the above expression are given in nuclear matter as the volume integral of
$V^{*d}(s,\rho)$ for the Hartree contributions, e.g.
\begin{equation}
    \tilde V_{qq'}^{d}(\rho) = \int d^{3}s \, V_{qq'}^{*d}(s, \rho),
\end{equation}
while the non-local Fock contributions are averaged over the Fermi spheres, e.g.
\begin{equation}
    \tilde V_{qq'}^{e}(\rho;k_{F_{q}},k_{F_{q'}}) = \int d^{3}s \,
    \rho_{SL}(sk_{F_{q}}) \rho_{SL}(sk_{F_{q'}}) \, V_{qq'}^{*e}(s,\rho)
\end{equation}
For Yukawa form factors the exchange integrals can be evaluated in closed form which,
however, is omitted here. In symmetric nuclear matter, Eq.(\ref{easym}) simplifies to
\cite{baduri,oertzen}
\begin{eqnarray} \label{e/a}
    \frac{E}{A}(\rho, \rho_{\tau}=0) & = & \frac{3}{5} \tau + \frac{1}{2} \,
    \rho \left[ \tilde V_{0}^{d}(\rho) + \tilde V_{0}^{e}(\rho,k_F) \right]
    \\ & = & \frac{3}{5} \tau + \frac{1}{2} \, \rho f_{0}(\rho) \left[ \int
    d^{3}s \, V_{0}^{d}(s) + \int d^{3}s \, \rho_{SL}^{2}(sk_{F}) \,
    V_{0}^{e}(s) \right].
\end{eqnarray}
where $k_{F}=(3/2 \pi^{2}\rho)^{1/3}$ and $\tau = \frac{\hbar^{2}
k_{F}^{2}}{2m}$. For given s$_{0,\tau}$ and exponent $\beta$ the
remaining parameters of f$_0$ are determined by
imposing the conditions for the nuclear matter binding energy, the saturation
density and the compressibility at the equilibrium density $\rho_0$,
\begin{eqnarray}
    & &\left. \frac{E}{A}(\rho, \rho_{\tau}=0)\right|_{\rho = \rho_{0}} =
    -a_{V} \\ & &P = \rho^{2}\, \left.
    \frac{d\left(E/A\right)}{d\rho}\right|_{\rho = \rho_{0}} = 0 \\ &
    &K_{\infty} = 9 \rho^{2}\, \left. \frac{d^{2} \left(E/A\right)}
    {d\rho^{2}}\right|_{\rho = \rho_{0}}
\end{eqnarray}
and the isovector vertex renormalization f$_\tau$ is obtained from
the symmetry energy $\varepsilon_{\tau}(\rho)$, defined by
\begin{equation} \label{modea}
    \frac{E(\rho, \rho_{\tau})}{A}  = \frac{E(\rho, 0)}{A} +
    \left(\frac{\rho_{\tau}}{\rho}\right)^{2} \varepsilon_{\tau}(\rho) +
    \left(\frac{\rho_{\tau}}{\rho}\right)^{4} \varepsilon_{4}(\rho) + \dots
\quad .
\end{equation}
At the saturation density $\rho=\rho_{0}$ the first term in Eq.(\ref{modea})
gives the binding energy per particle of symmetric nuclear matter, $-a_{V}$, and
$\varepsilon_\tau$ is the symmetry energy coefficient
\begin{equation} \label{asym}
    a_{s} = \rho_{0}^{2} \left. \frac{\partial^{2}
    E(\rho_{0}, \rho_{\tau})}
	    {\partial \rho_{\tau}^{2}} \right|_{\rho_{\tau} = 0}
\end{equation}
which is known to range between 28~MeV and 32~MeV \cite{baduri,lyon}.

For a finite range interaction the above procedure for fixing the isovector
properties of asymmetric nuclear matter involves higher order contributions from
the intrinsic momentum structure of the interaction. They appear because the
asymmetry in the proton and neutron Fermi momenta induces an effective isospin
symmetry breaking component via the exchange integrals extending over the two
Fermi spheres. Numerically, the mixing of the isoscalar and the isovector channel
is noticed immediately from the separate averaging of the exchange contributions
over the proton and neutron Slater densities $\rho_{SL}(k_{Fp})$ and
$\rho_{SL}(k_{Fn})$, Eq.(\ref{e/a}). In the direct part or for a momentum
independent contact interaction, respectively,	these contributions are absent.
In those cases the potential energy can always be written in the form $\rho^{2}
V_{0} + \rho_{\tau}^{2} V_{\tau}$, which separates exactly the isoscalar and the
isovector part of the interaction (except for the case that $V_0$ would depend
intrinsically also on $\rho_\tau$). Skyrme interactions account to some extent
for the isovector exchange effect by the momentum dependent components.	 At
extreme high asymmetries, however, differences to a finite range interaction
might become noticeable.

\begin{table}
\caption{Parameters of the unrenormalized M3Y interaction in the odd-even (first block),
    the spin-isospin (second block), the p-n representation as defined
    (third block) and the effective spin-isospin representation
    (fourth block) as defined in App. \ref{app}.}
\begin{tabular}{|l||r|r|r||r|}
$r_{0}$ [fm] & 0.250 & 0.400 & 1.414 & Volume integral\\ $m=\frac{\hbar
c}{r_{0}}$ [MeV] & 780 & 490 & 140 & [MeV f$m^{3}$]\\
\hline
$V_{SE}$ & 11466.00 & -3556.00 & -10.46 &  -980.28\\ $V_{TE}$ & 13967.00 &
-4594.00 & -10.46 & -1324.02\\ $V_{SO}$ & -1418.00 &   950.00 &	 31.39 &
1600.77\\ $V_{TO}$ & 11345.00 & -1900.00 &   3.49 &   823.43\\
\hline
$V_{0}$	     & 11062.00 & -2538.00 & 0.00 & 131.17\\ $V_{\sigma}$     &
939.00 &   -36.00 & 0.00 & 155.40\\ $V_{\tau}$	     &	 314.00 &   224.00 & 0.00
& 241.33\\ $V_{\sigma\tau}$ &  -969.00 &   450.00 & 3.49 & 295.53\\
\hline
$V_{pp}^{d}$ & 11375.00 & -2314.00 &   0.00 &	372.51\\ $V_{pn}^{d}$ & 10748.00
& -2761.00 &   0.00 &  -110.16\\ $V_{pp}^{e}$ & -5642.00 &   536.00 &  -5.23 &
-862.65\\ $V_{pn}^{e}$ &  2594.00 & -1574.00 & -10.46 &	 -1127.93\\
\hline
$V_{0}^{d}$    & 11062.00 & -2538.00 &	0.00 &	131.17\\ $V_{\tau}^{d}$ &
314.00 &   224.00 &  0.00 &  241.33\\ $V_{0}^{e}$    & -1524.00 &  -519.00 &
-7.85 & -995.29\\ $V_{\tau}^{e}$ & -4118.00 &  1055.00 &  2.62 &  132.64\\
\end{tabular}
\label{tm3y}
\end{table}

%-------------------------------------------------------------------------------------
\subsection{Results for Infinite Nuclear Matter}

The isoscalar vertex function f$_{0}(\rho)$ was adjusted to the empirical values
of the binding energy ($a_{V} \simeq 16$ MeV), the saturation density
($\rho_{0}\simeq$0.159~fm$^{-3}$) and the compressibility K$_\infty$.
Empirically, K$_\infty$ is not well known (values range from 200~MeV to 350~MeV),
but recent studies, e.g. \cite{lyon}, converge to a value of about
K$_\infty$=230~MeV. For the symmetry energy coefficient serving to fix f$_\tau$
$a_{s}=\varepsilon_{\tau}(\rho_{0}$=0.159~fm$^{-3}$)=32~MeV is used.

In Tab. \ref{f0} two sets of vertex renormalization parameters,
Eq.(\ref{vertex_fit}) are displayed. The first set was obtained with s$_{0}=1$,
i.e. leaving the free space interaction unchanged, whereas the second set with
s$_{0}=5$ renormalizes the asymptotic behavior of the interaction as well. This
value of s$_{0}$ was determined from the binding energies of finite nuclei as
will be discussed in the next section.

\begin{table}
\caption{Parameterization of f$_{0}(\rho)$ and f$_{\tau}(\rho)$}
\begin{tabular}{crrrrrr}
& $a_1^0$ & $a_2^0$ & $a_3^0$ & $\rho_{0}\,[\mbox{fm}^{-3}]$ & $K$ [MeV] &
$\left(E_{B}/A\right)$ [MeV] \\
\hline
s$_{0} = 1$ &  1.365 & -1.839 &	 0.379 & 0.159 & 230 & -16.00 \\ s$_{0} = 5$ &
-2.127 &  2.032 & -0.724 & 0.159 & 230 & -16.00 \\
\hline
& $a_1^{\tau}$ & & & & & $\varepsilon_{\tau}(\rho_{0})[MeV]$ \\
\hline
s$_{\tau} = 1$ & .1688 & & & & & 32.00\\
\end{tabular}
\label{f0} \label{sfit} \label{ft} \label{tfit}
\end{table}

\begin{figure}
\begin{center}
\epsfig{file=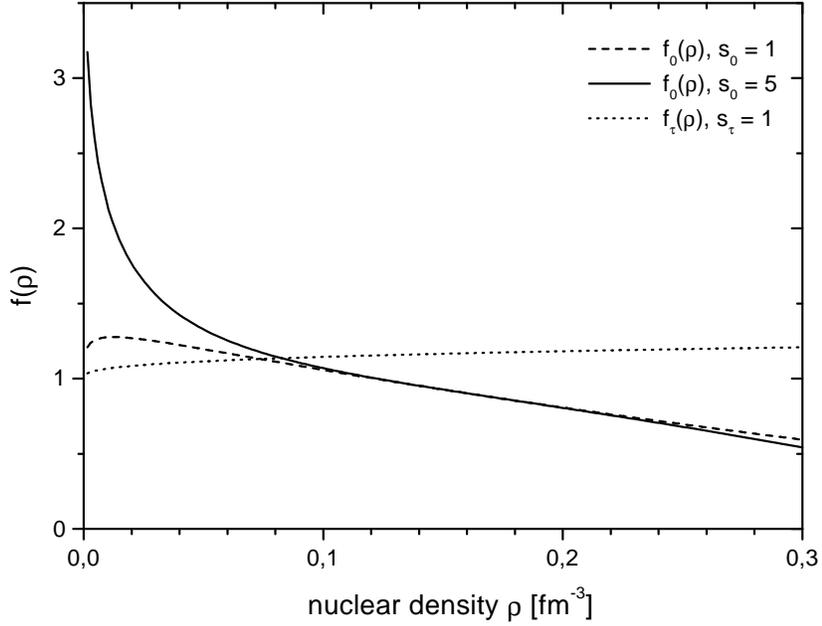, width=14cm}
\caption{Density dependent isoscalar f$_{0}$
(s$_{0} = 1$ dashed line, s$_{0} = 5$ solid line) and isovector f$_{\tau}$
(dotted line) vertex functions of the D3Y interaction.}
\label{vertex}
\end{center}
\end{figure}

\begin{figure}
\begin{center}
\epsfig{file=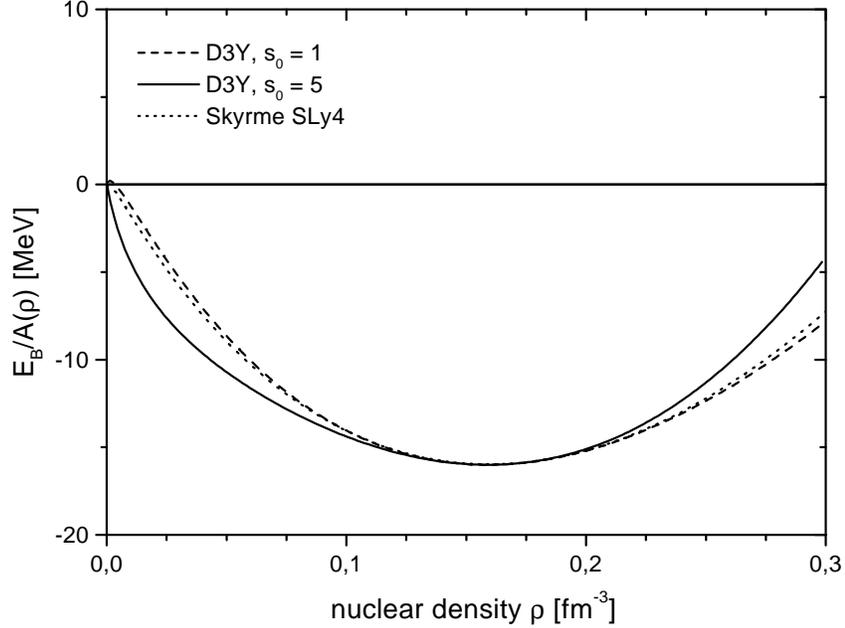, width=14cm}
\caption{
Nuclear matter equation of state calculated with the density dependent D3Y
interaction for the parameterizations s$_{0}=1$ (dashed line) and s$_{0}=5$
(solid line) and the Skyrme SLy4 interaction (dotted line). }
\label{eos}
\end{center}
\end{figure}

Fig. \ref{vertex} shows the density dependence of the vertex correction
functions f$_{0}$ and f$_{\tau}$. In the isovector channel only a rather
small deviation from unity is found, and f$_{\tau}$ increases only
slightly for higher densities. This reflects the fact that the original
interaction already has a symmetry coefficient close to 32 MeV. In the
isoscalar channel one realizes an increase of the interaction at low
densities and a suppression of the interaction at higher densities. The
low density behavior obviously scales with s$_{0}$. In Fig. \ref{eos}
the equations of state (EoS) for the D3Y and  the Skyrme SLy4
interaction \cite{lyon} are compared. For s$_0$=1 the D3Y and the SLy4
results are in very close agreement over the full range of densities.
According to the behavior of f$_{0}$, the binding energy increases at
densities lower and decreases at densities higher than the saturation
density for s$_{0}>$1 while retaining its value at the equilibrium
density.

\begin{figure}
\begin{center}
\epsfig{file=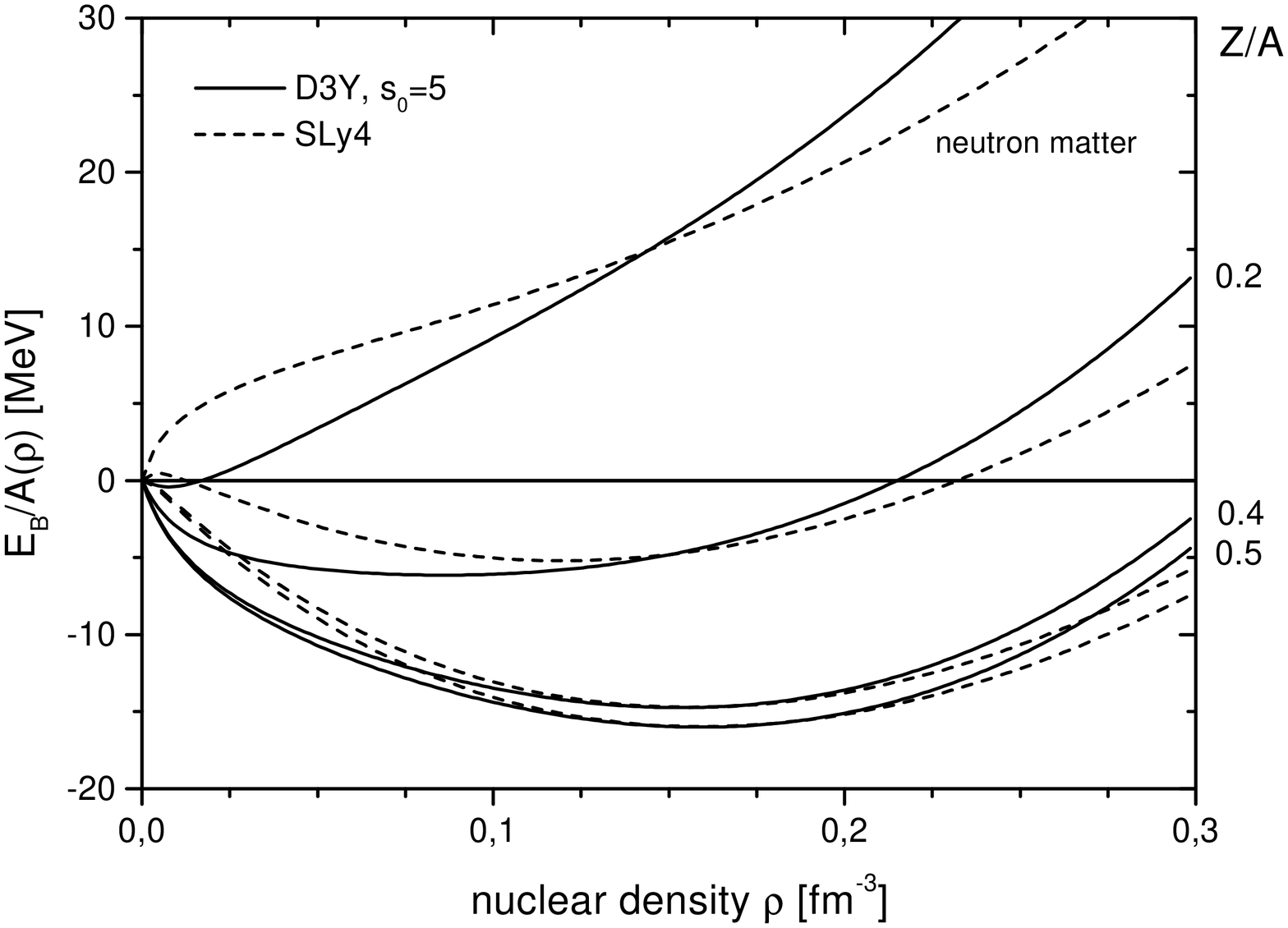, width=14cm}
\caption{
Equation of state for asymmetric nuclear matter calculated with the density
dependent D3Y interaction for the parameterizations s$_{0} = 5$ (solid line) and
the Skyrme SLy4 (dashed line). Shown is the EoS for neutron matter and for the
asymmetry ratios $Z/A = 0.2, 0.4, 0.5$. }
\label{eos_asym}
\end{center}
\end{figure}

The isovector properties of the D3Y interactions are illustrated in
Fig. \ref{eos_asym}, where the EoS for asymmetric nuclear matter is
displayed for different asymmetry ratios $Z/A =
\rho_{p}/ \rho$. Stable nuclei typically cover the range $Z/A = 0.5
\dots 0.4$ (e.g. $^{40}$Ca, $^{208}$Pb). In exotic nuclei the new
regions with $Z/A = 0.4 \dots 0.3$ become accessible and far of stability
$Z/A=0.2$ may be reached (e.g $^{10}$He). It is remarkable that neutron matter is
still bound but this occurs at extremely low densities. The reason for this
behavior is the strong attraction of the D3Y with s$_{0}=5$ at very low densities
and, as such, is not very conclusive. In stable nuclei these density regions will
rarely contribute to the total binding. However, a different situation may be
expected far off stability, especially in halo nuclei, where the binding
sensitively depends on the interactions at low density. Comparing the D3Y to the
SLy4 interaction one sees that both interactions are in agreement around the
saturation density for asymmetry ratios of $Z/A = 0.5 \dots 0.4$. The differences
showing up at low densities and large asymmetries are mainly due to the stronger
attraction of the D3Y isoscalar interaction with s$_0$=5.

\begin{figure}
\begin{center}
\epsfig{file=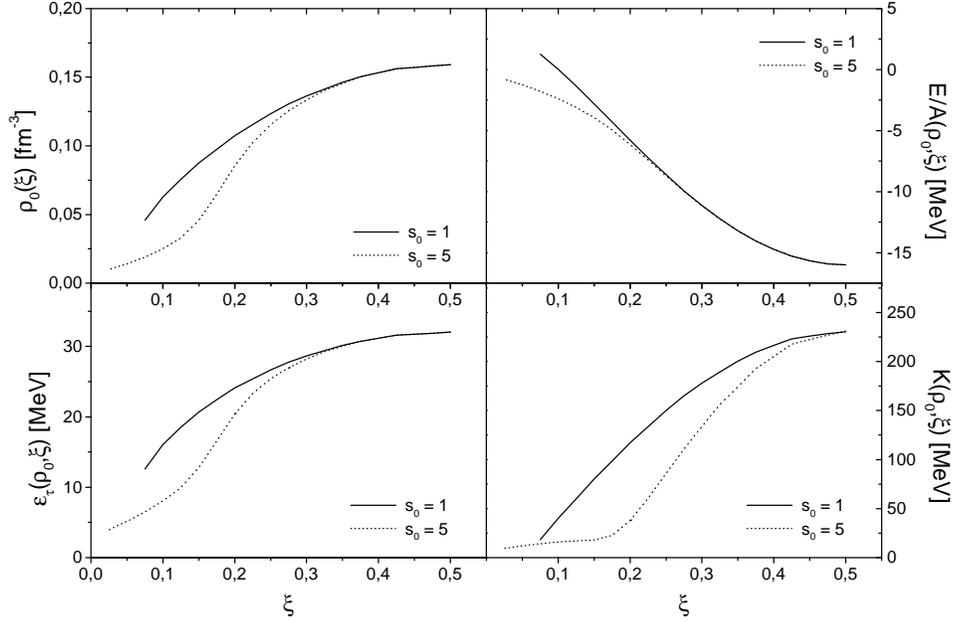, width=14cm}
\caption{
Equilibrium density $\rho_{0}$ of the EoS (upper left), binding energy per
nucleon $E_{B}/A$ at $\rho_{0}$ (upper right), symmetry coefficient
$\epsilon_{\tau}$ at $\rho_{0}$ (lower left) and the compressibility
K$_\infty$($\rho_{0}$) (lower right ), all plotted as a function of the neutron
excess or asymmetry ratio $\xi=Z/A$. Results are shown for the D3Y interaction
with s$_{0} = 1$ and s$_{0} = 5$. }
\label{asymrat}
\end{center}
\end{figure}

In Fig. \ref{asymrat} the EoS for different asymmetry ratios and the two sets of
parameters are examined. The equilibrium density $\rho_{0}(\xi)$, the energy per
nucleon $E_{B}(\xi)/A$, the symmetrie energy $\epsilon_{\tau}(\xi)$ and the
compressibility K$_\infty$($\xi$) for $\xi$=Z/A are shown. A higher value of
s$_{0}$ is seen to lead to a lower equilibrium density for strong asymmetries. In
addition the binding energy at the equilibrium density is higher due to the
stronger attraction of this parameterization of the interaction at low densities.
The lower left part shows the symmetry energy of the EoS in dependence of the
proton excess. One realizes a decrease of the symmetry energy for strong
asymmetries and a higher value of s$_{0}$ corresponding to the shift of the
saturation density to lower densities. The compressibility approaches zero at
high asymmetries with a remarkable flat slope.

The properties of the D3Y interaction in the particle-particle channel were
investigated by considering like--particle pairing in symmetric nuclear matter.
Pairing, being mediated by the singlet-even part $V_{SE}$ of the interaction, was
examined by Hartree-Fock-Bogoliubov (HFB) calculations. The in--medium SE
particle--particle interaction was obtained by back transformation from the
renormalized particle--hole interactions using the representation in terms of the
exchange interactions (see App. \ref{app}), leading to $ V_{SE}^{*}(\rho) = 2
\sum_{\gamma=0,\tau}\left[ V_{\gamma}^{*d}(\rho) + V_{\gamma}^{*e}(\rho)\right] $.

As in refs. \cite{kutcha1,kutcha2} the following set of equations was solved
simultaneously:
\begin{eqnarray} \label{hfb}
    \rho(k_{F}) & = & 4 \int \frac{d^{3}k}{(2 \pi)^{3}} v^{2}(k,k_{F}) \\
    v^{2}(k,k_{F}) & = & \frac{1}{2} \left\{ 1 - \frac{\epsilon(k,k_{F}) -
    \lambda} {\sqrt{(\epsilon(k,k_{F}) - \lambda)^{2} + \Delta^{2}(k,k_{F})}}
    \right\} = 1 - u^{2}(k,k_{F}) \\ \epsilon(k,k_{F}) & = &
    \frac{\hbar^{2}k^{2}}{2m} + 4 \int \frac{d^{3}k'}{(2 \pi)^{3}} \langle k k' |
    V^{*} | k k' - k' k \rangle v^{2}(k',k_{F}) \\ \Delta(k,k_{F}) & = & - \int
    \frac{d^{3}k'}{(2 \pi)^{3}}
	    \langle k -k | V_{SE}^{*} | k' -k' \rangle u(k',k_{F})
	     v(k',k_{F})
\end{eqnarray}

\begin{figure}
\begin{center}
\epsfig{file=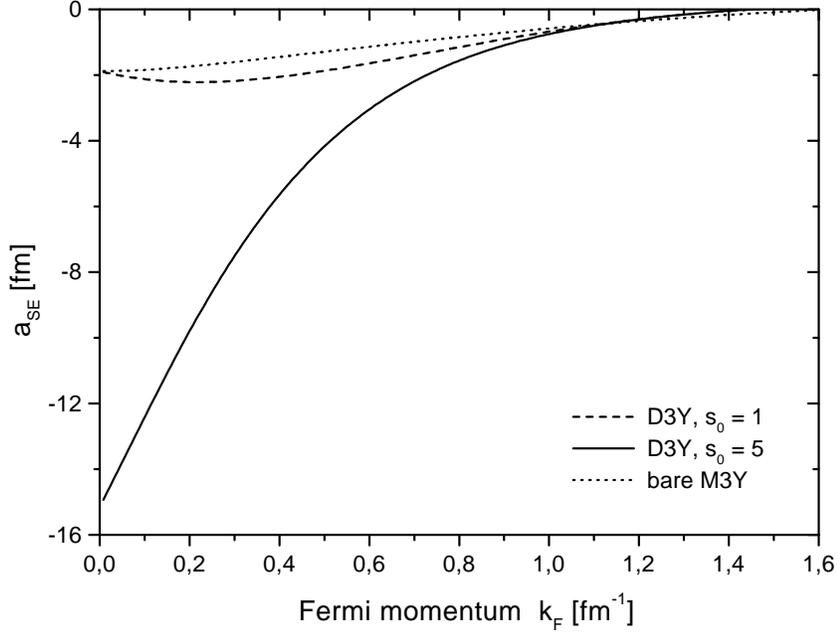, width=14cm}
\caption{
Scattering length $a_{SE}^{pp}(k_{F})$ for relative s--wave states. Results are
shown for the bare M3Y (dotted) and the D3Y with the parameterizations s$_{0}
= 1$ (dashed line) and s$_{0} = 5$ (solid line). }
\label{vse}
\end{center}
\end{figure}

\begin{figure}
\begin{center}
\epsfig{file=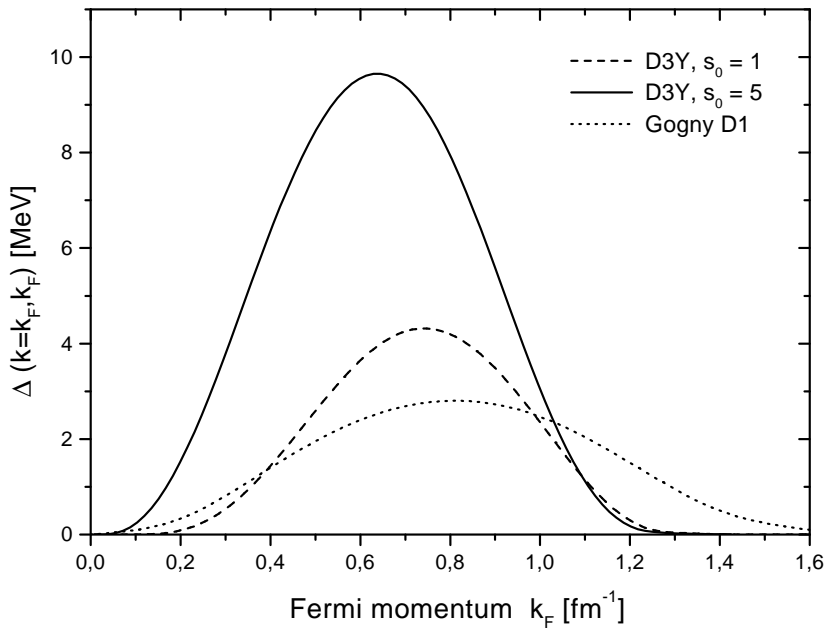, width=14cm}
\caption{
Gap at the Fermi momentum in nuclear matter. Results for the D3Y interaction with
s$_{0} = 1$ (dotted line) and s$_{0} = 5$ (solid line) and the Gogny D1 (dashed
line) are shown. }
\label{gap}
\end{center}
\end{figure}

Fig. \ref{vse} shows the calculated in--medium scattering length for like
particles (pp/nn) $a_{SE}^{pp}(k_{F})$ at a given Fermi momentum $k_{F}$ for
different parameterizations of the interaction. Because of the density dependence
in the isovector and in the isoscalar channel the strength of the interaction
increases at low to intermediate densities. For the D3Y with s$_{0}=5$ the
pairing interaction is enhanced by a factor of about 8 at low densities over the
values of the original M3Y because of the asymptotic scaling of the isoscalar
interaction. This corresponds to about 90\% of the s-wave strength of the free NN
T-matrix with a scattering lenght of $a_{SE}^{pp}=-17.1$fm and leads to higher
values of the gap in nuclear matter. In Fig. \ref{gap} $\Delta(k,k_{F})$ at
k=k$_F$ is shown. For the bare M3Y the position of the maximum gap lies at about
k$_{F}\simeq$0.8~fm$^{-1}$. About the same result is obtained in HFB calculations
with the Gogny force \cite{kutcha1,kutcha2} and for several Skyrme interactions
\cite{takahara}. The position of the maximum gap in the D3Y--calculations is
shifted to slightly lower densities with $k_{F}
\simeq$=0.75~fm$^{-1}$ resp. k$_{F} \simeq$0.63~fm$^{-1}$ for s$_{0}$=1 resp.
s$_{0}$=5. One finds that the height of the maximum increases by lowering the
position of the maximum. Our results of $\Delta(k_{F},k_{F}) \simeq 2$ MeV resp.
$\Delta(k_{F},k_{F}) \simeq 4$ MeV for the M3Y and the D3Y with s$_{0}=1$ are in
agreement with the Gogny force that yields a value of $\Delta(k_{F},k_{F}) \simeq
3$ MeV \cite{kutcha1,kutcha2}. The calculated maximum gap for the D3Y with
s$_{0}=5$ has a much higher value of about 9.5 MeV. The reason for this is the
much stronger pairing interaction, especially at lower densities, as seen in Fig.
\ref{vse}. Comparing the value of the gap at the saturation density $k_{F}=1.33
fm^{-1}$ with other calculations we find a value of about 0.2 MeV for the M3Y and
a nearly vanishing value for the D3Y interaction. For the Gogny force one finds a
value of about 0.6 MeV but most other calculations favor a small to vanishing gap
\cite{takahara,chen,kennedy}.

%-------------------------------------------------------------------------------------
\section{Hartree--Fock Calculations for Finite Nuclei} \label{4}

The density dependent D3Y interaction derived in the preceding section for
nuclear matter is applied in HF--calculations for finite nuclei. Stable nuclei
close to the N$\simeq$Z line are chosen first. In these calculations mainly the
isoscalar properties of the interaction are investigated. A more general test,
including also isovector dynamics, is obtained from calculations for even-even
Sn--isotopes covering the range from the proton dripline ($^{100}$Sn) to the
A=140 mass region. The calculations for stable and unstable nulcei are performed
without readjustment of parameters from their values derived for nuclear matter.

The general structure of the HF--equations for a density dependent NN-interaction in the DME
approach was already derived in Sect. \ref{hfe}. In the practical applications the DME is
used only for the exchange contributions to the HF potentials. Especially at higher density
and thus Fermi momentum exchange is dominated by the short range pieces of the interaction
for which the local momentum approximation underlying the DME is most appropriate. Different
from the original work of Negele and Vautherin \cite{negvaua} the direct, i.e. Hartree
potentials are treated in full finite range.

The D3Y density dependence was introduced as a local renormalization of the
vertices which requires to replace the interaction by \cite{LF2}
\begin{equation}\label{vdir}
	V^{*d}_{\gamma}({\bf r}_1,{\bf r}_2; \rho)= g_{\gamma}({\bf
	r}_1)V^d_{\gamma}({\bf r}_1-{\bf r}_2)g_{\gamma}({\bf r}_2)
\end{equation}
with $g^2_{\gamma} = f_{0,\tau}$ and where $g_{\gamma}({\bf r}) =
g_{\gamma}(\rho({\bf r}))$ are the in--medium isoscalar and isovector vertex
renormalizations, respectively. $V^d_{\gamma}({\bf r}_1-{\bf r}_2)$ denotes the
unrenormalized ''bare'' interaction in these channels. This ensures a reliable
treatment of the density dependence also in the surface region where the vertices
vary rapidly (see Fig. \ref{vertex}). The direct potentials are obtained as in
Eq.(\ref{sppd}) by folding the proton and neutron densities, respectively, with
the corresponding -- now density dependent -- direct interactions. To accomplish
this we express the interactions $V_{qq'}^{d}({\bf r}_{1},{\bf r}_{2})$ by
$V^{*d}_{\gamma}({\bf r}_1,{\bf r}_2; \rho)$. With the DME the exchange or Fock
potentials are replaced by local potentials whose strengths are defined in
Eq.(\ref{delta}). In order to preserve this separation of the center--of--mass
and relative coordinates ${\bf r}$ and ${\bf s}$ the interaction must also be
separable in ${\bf r}$ and ${\bf s}$. Assuming that the resulting error is small
compared to the error introduced by the DME we therefore replace in the exchange
interaction, as a first approximation, the individual coordinates of the
interacting particles by their center--off--mass coordinate. This leads to
\begin{equation}\label{vexch}
	V^{*e}_{\gamma}({\bf s},{\bf r}; \rho)= g^{2}_{\gamma}({\bf r})
	V^e_{\gamma}({\bf s})
\end{equation}
by which we express the exchange interactions $V_{qq'}^{e}({\bf s})$.

%-------------------------------------------------------------------------------------
\subsection{Details Of Numerical Calculation}
For our calculations we only consider the stationary ground state of spherical
even-even nuclei, therefore the single particle wave functions can be separated
as
\begin{equation} \label{ewave}
	\phi_{k_{q}}({\bf r}) =	 \frac{R_{\alpha_{q}}(r)}{r} \Psi_{ljm}(\theta,
	\phi)
\end{equation}
The functions $\Psi_{ljm}(\theta, \phi)$ are spinor spherical harmonics
\cite{edmonds}, the index $\alpha_{q}$ represents the quantum numbers
$\alpha_{q}=\{n,l,j\}$ for protons and neutrons, resp. This allows us to simplify
the density and the kinetic energy density to
\begin{eqnarray}
	\rho_{q}(r) & = & \frac{1}{4 \pi r^{2}} \,\sum_{\alpha_{q}}\,
	v_{\alpha_{q}}^{2}(2 j_{\alpha_{q}} + 1) R_{\alpha_{q}}^{2}(r) \\
	\tau_{q}(r) & = & \frac{1}{4 \pi} \, \sum_{\alpha_{q}}\,
	v_{\alpha_{q}}^{2} (2 j_{\alpha_{q}} + 1) \left[ \left( \partial_{r}
	\frac{R_{\alpha_{q}}(r)}{r} \right)^{2} + \frac{ l_{\alpha_{q}}
	(l_{\alpha_{q}}+1)}{r^{2}} \,
	\left(\frac{R_{\alpha_{q}}(r)}{r}\right)^{2} \right]
\end{eqnarray}
The occupation weights $v_{\alpha_{q}}^{2}$ for a state $\alpha_{q}$ are
determined by pairing correlations. Since it was found that pairing gives minor
to negligible contributions to the nuclei considered here, the BCS approximation
was used. A constant pairing matrix element of $G_{q} = 23$ MeV/nucleon was used
and the standard set of BCS equations \cite{baduri} was solved independently for
protons and neutrons, respectively.

For protons also the Coulomb interaction contributes to the Hartree--Fock
potential $U^{HF}$:
\begin{equation}
	U_{Coul}({\bf r}) = \int d^{3}r_{1} \, \rho_{p}({\bf r}_{1}) \,
	\frac{e^{2}}{|{\bf r} - {\bf r}_{1}|} \,\,\,\,-\,\,\,\, \left(
	\frac{3}{\pi} \right)^{1/3} e^{2} \rho_{p}({\bf r})^{1/3}
\end{equation}
The Coulomb exchange term is treated in the local density approximation rather
than in the full DME because its contribution to the potential energy is small.

The spin-orbit potential is treated as in \cite{negvaua,vaubrink}, leading to a
contribution to the potential energy $E_{s.o.} = \int d^{3}r\, H_{s.o.}({\bf r})$
with
\begin{eqnarray}
	H_{s.o.}({\bf r}) & = & - \frac{1}{2} W_{0} \left\{ \rho({\bf r}) \, {\bf
	\nabla} {\bf J}({\bf r}) + \sum_{q} \rho_{q}({\bf r}) \, {\bf \nabla}
	{\bf J}_{q}({\bf r}) \right\} . \nonumber
\end{eqnarray}
The spin-orbit strength $W_{0}$ is related to the two-body spin-orbit potential
in the short--range limit by
\begin{equation} \label{w0}
	W_{0} = -\frac{2 \pi}{3} \int V_{LS}(s) s^{4} \, ds
\end{equation}
Using the M3Y parameterization \cite{toki} one finds a value of $W_{0}
\sim 105$ MeV fm$^5$ but in correspondence to most Skyrme interactions
$W_{0} = 120$ MeV fm$^5$ from reference \cite{vaubrink} is chosen. After
variation this term gives a contribution
\begin{equation}
	U_{q}^{s.o.}({\bf r}) = -\frac{1}{2} W_{0} \left[ \rho({\bf r}) \, {\bf
	\nabla} {\bf J}({\bf r}) + \rho_{q}({\bf r}) \, {\bf \nabla} {\bf
	J}_{q}({\bf r})\right],
\end{equation}
to the Hartree--Fock potential and a single particle spin-orbit potential
\begin{equation}
	U_{q}^{{\bf l} {\bf s}}(r) = \frac{1}{2} W_{0} \frac{1}{r}\,
	\frac{d}{dr}\, \left[\rho(r) + \rho_{q}(r)  \right]  {\bf l} \cdot {\bf
	s}
\end{equation}
where the spin density is defined as
\begin{equation}
	J_{q}(r) = \frac{1}{4 \pi r^{3}} \, \sum_{ \alpha_{q}} \, (2
	j_{\alpha_{q}} + 1) \left[ j_{\alpha_{q}} (j_{\alpha_{q}} + 1) -
	l_{\alpha_{q}}(l_{\alpha_{q}} + 1) - \frac{3}{4} \right]
	R_{\alpha_{q}}^{2}(r).
\end{equation}
%-------------------------------------------------------------------------------------
\subsection{Results for Stable Nuclei}

\begin{figure}
\begin{center}
\epsfig{file=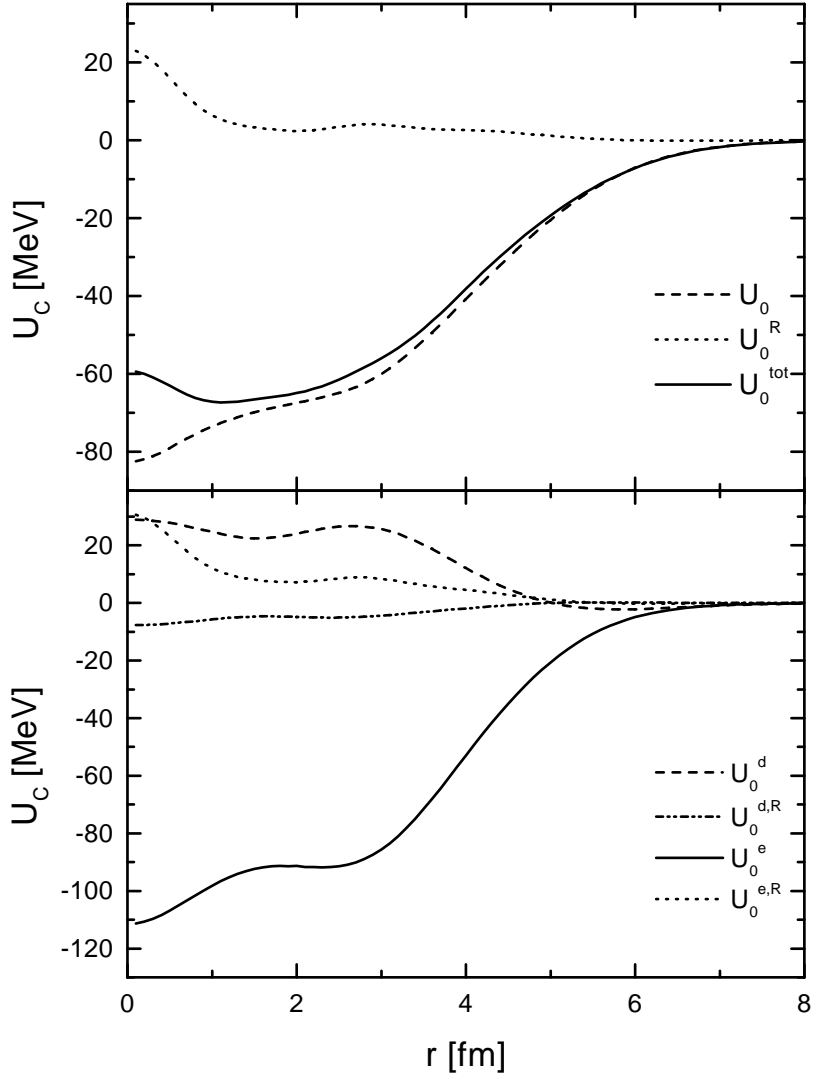, width=12cm}
\caption{
Single particle potential for ${}^{40}$Ca calculated with the D3Y with $s_{0}=5$.
In the lower part, contributions to the total isoscalar potential from the direct
$U_{0}^{d}$ and the exchange term $U_{0}^{e}$ of the normal Hartree-Fock
potential as well as the rearrangement potentials ($U_{0}^{d,R}$, $U_{0}^{e,R}$)
are shown. In the upper part, the total isoscalar potential $U_{0}^{tot}$ as well
as its contributions, the attractive HF potential $U_{0}$ and the repulsive
rearrangement potential $U_{0}^{R}$, are shown. }
\label{rearr}
\end{center}
\end{figure}

The various contributions to the nuclear part of the self--consistent DME
mean--field in $^{40}$Ca are shown in Fig. \ref{rearr}. The isoscalar Hartree
potential is weak and repulsive and binding is essentially mediated by the Fock
parts. The Fock potential is strongly attractive with a depth at the center of
about -110~MeV. The sum of the Hartree and Fock terms leads to an average
potential depth of -70~MeV. The direct and exchange rearrangement potentials
reduce the depth especially close to the center. The mean--field agrees
qualitatively with other non--relativistic HF results, e.g \cite{vaubrink}. It
also resembles closely the Schroedinger-equivalent potential from a relativistic
density dependent Hartree calculation \cite{LF1,LF2} except for a more pronounced
shell structure.

Of special interest is the influence of the rearrangement potential whose
influence has proven to be important as found in comparable non-relativistic
\cite{vaubrink} and relativistic calculations \cite{LF1,LF2}. Rearrangement is of
importance for the single particle potential and the energy spectrum. Because the
rearrangement potential is always repulsive it lowers the single particle
potential and leads to a change in the single particle spectra and the charge
radii. In the calculation of the total binding energy the rearrangement effects
cancel out. Taking this into account the total binding energy is given as
\begin{equation}
	E_{B} = \sum_{i=1}^{A} \epsilon_{i} - \frac{1}{2} \sum_{q=p,n}\int d^{3}
	r \rho_{q}(r) U^{HF}_{q}(r) - \sum_{q=p,n} E_{q}^{R} + E_{pair},
\end{equation}
with the rearrangement energy calculated from Eq.(\ref{emass}) and
Eq.(\ref{rpot}):
\begin{equation}
	E_{q}^{R} = \int d^{3}r\, \rho_{q}(r)\, U^{R}_{q}(r)
		  + \int d^{3}r\, \tau_{q}(r)\,
		  \frac{\hbar^{2}}{2}\left(\frac{1}{m_{q}^{*}(r)}-\frac{1}{m}
		  \right)
\end{equation}

\begin{figure}
\begin{center}
\epsfig{file=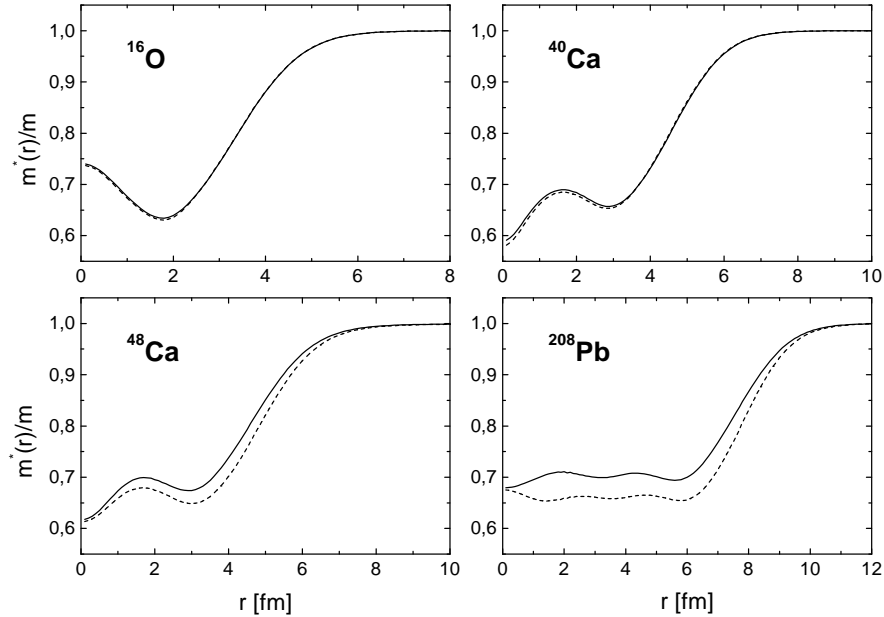, width=14cm}
\caption{
Effective mass of protons (solid line) and neutrons (dashed line) in closed shell
nuclei calculated with the density dependent D3Y interaction (with the
parameterization $s_{0} = 5$.) }
\label{emass_plt}
\end{center}
\end{figure}

An important contribution to mean--field dynamics is provided by the momentum
dependence of the interaction. Since the exchange parts of the D3Y interaction
were found to be most important for nuclear binding the question arises whether
the momentum structure is described realistically. The effective mass m$^*_q$,
Eq.(\ref{emass}) is completely determined by the Fock terms, since they are the
only source of non--localities in the present model. From Fig. \ref{emass_plt} it
is seen that a very reasonable value of about $m^*_q/m\simeq$0.68 is obtained in
the interior of a heavy nucleus like $^{208}$Pb which closely agrees with other
theoretical results and determinations from proton--nucleus elastic scattering
\cite{mahaux}. It is worthwhile to emphasize that this result was obtained
without readjustments of parameters. The D3Y interaction (or better to say the
original M3Y G--matrix \cite{toki}) has apparently a realistic mixture of short
and long range contributions accounting reasonably well for the intrinsic
momentum structure of an in--medium HF interaction. The results of this section
also confirm the DME approach. Especially promising results are obtained in
conjunction with the effective ''Fermi momentum'' q$_F$, Eq.(\ref{modfermi}), by
which non--local effects are accessible in a self--consistent way.

\begin{table}
\caption{
Charge radii $r_{C}$ [fm] for closed shell nuclei for different parameterizations
of the D3Y interaction}
\begin{tabular}{lcccc}
& ${}^{16}$O & ${}^{40}$Ca & ${}^{48}$Ca & ${}^{208}$Pb \\
\hline
$s_{0} = 1$ & 2.79 & 3.51 & 3.52 & 5.51 \\ $s_{0} = 5$ & 2.88 & 3.61 & 3.60 &
5.59 \\
\hline
exp.	    & 2.73 & 3.49 & 3.47 & 5.50 \\
\end{tabular}
\label{magicrc}
\end{table}

\begin{figure}
\begin{center}
\epsfig{file=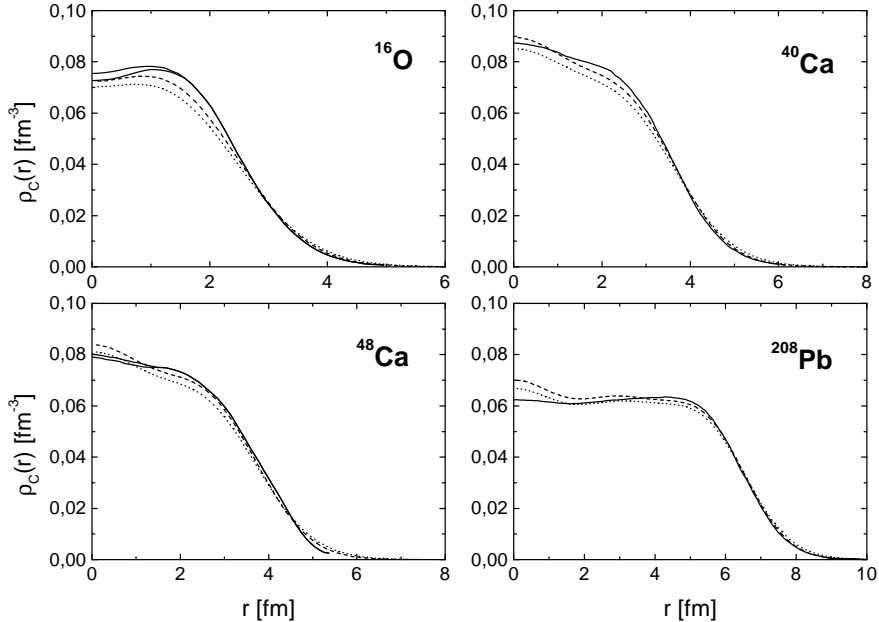, width=14cm}
\caption{
Charge density distributions of closed shell nuclei from density dependent
Hartree-Fock calculations using the D3Y interaction. Results for $s_{0} = 1$
(dashed line) and $s_{0} = 5$ (dotted line) are shown. Experimental densities and
their uncertainties are denoted by solid lines (taken from Ref.
\protect\cite{grange})
.}
\label{density}
\end{center}
\end{figure}

Fig. \ref{density} compares the calculated charge density distributions of the
above closed--shell nuclei with experimental data from elastic electron
scattering \cite{grange}. The theoretical point particle distribution was folded
with a Gaussian proton form factor \cite{vaubrink} with $\sqrt{\langle r^{2}
\rangle_{p}} = 0.8fm$. Tab. \ref{magicrc} shows the calculated charge radii in
comparison to the experimental data (taken from \cite{LF2}).

\begin{figure}
\begin{center}
\epsfig{file=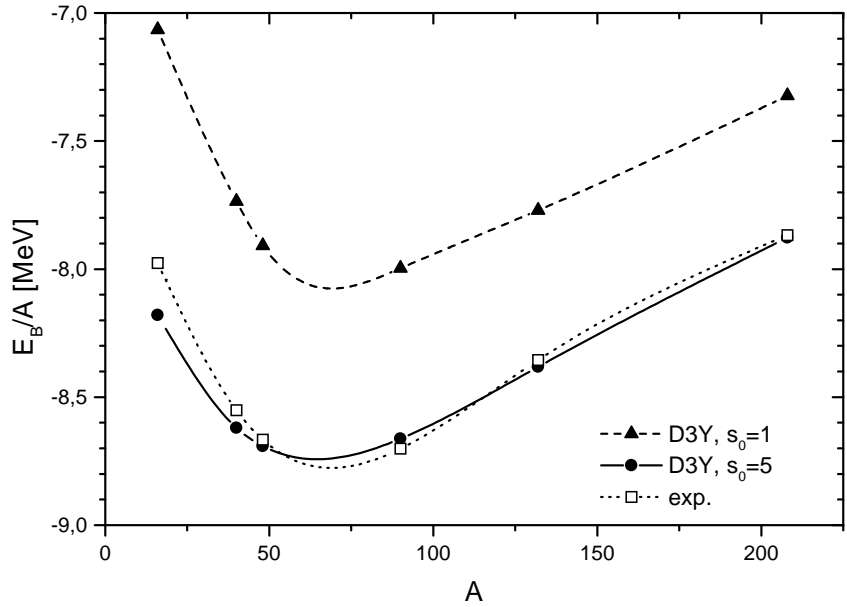, width=14cm}
\caption{
Binding energy per nucleon for closed shell nuclei from density dependent
Hartree-Fock calculations using the D3Y interaction. Results for $s_{0} = 1$
(triangles) and $s_{0} = 5$ (circles) are shown. Experimental binding energies
are denoted by open squares (taken from Ref.
\protect\cite{nndc}). Binding energies are shown for the nuclei $^{16}$O, $^{40}$Ca,
$^{48}$Ca, $^{90}$Zr, $^{132}$Sn and $^{208}$Pb. Lines are drawn to guide the
eye.}
\label{magic}
\end{center}
\end{figure}

The charge density in the nuclear interior is slightly overestimated for heavy
nuclei like $^{208}$Pb and slightly underestimated for the lighter nuclei. In all
cases, the density at large radii is higher than the experimental data which
leads to a overestimation of charge radii as can be seen in Tab. \ref{magicrc}.
In general the agreement with the experimental charge radii is satisfying and is,
except for $^{16}$O, $\leq 2\%$ for the D3Y with $s_{0}=1$ and $\leq 4\%$ for the
D3Y with $s_{0}=5$. In contrast to this, the binding energies calculated with
$s_{0}=1$ are strongly underestimated by more than 0.5 MeV per nucleon, whereas
the D3Y with $s_{0}=5$ reproduces the experimental data \cite{nndc} very well
(Tab. \ref{magicea}). One sees that an improvement in the binding energies leads
to higher values of the charge radii. In general it is not possible to reproduce
simultaneously both the binding energies and the charge radii with one parameter
set. This is in agreement with calculations in the relativistic density dependent
field theory which typically give comparable results \cite{tHM,BM}. But generally
the agreement with the experimental binding energies of the D3Y with $s_{0}=5$ is
surprisingly good as can be seen in Fig. \ref{magic} realizing that the number of
used parameters is lower than the one in Gogny or Skyrme interactions.

The observed behavior of the D3Y for different parameterizations shows the
importance of the density dependence of the interaction. Although both
parameterizations have the same properties in the range of the saturation density
in nuclear matter the results for finite nuclei are quite different. This
behavior is caused by the differences at low densities. In addition, the
rearrangement terms enhance the effect of the density dependence.

\begin{table}
\caption{
Binding energy per nucleon $E_{B}/A$ [MeV] for closed shell nuclei for different
parameterizations of the D3Y interaction}
\begin{tabular}{lcccccc}
& ${}^{16}$O & ${}^{40}$Ca & ${}^{48}$Ca & ${}^{90}$Zr & ${}^{132}$Sn &
${}^{208}$Pb \\
\hline
$s_{0} = 1$ & -7.07 & -7.74 & -7.91 & -8.00 & -7.77 & -7.32 \\ $s_{0} = 5$ &
-8.18 & -8.62 & -8.69 & -8.66 & -8.38 & -7.88 \\
\hline
exp.	    & -7.98 & -8.55 & -8.67 & -8.70 & -8.36 & -7.87 \\
\end{tabular}
\label{magicea}
\end{table}

%-------------------------------------------------------------------------------------
\subsection{The Ground States of Sn Isotopes}
The Sn isotopes are of particular interest for nuclear structure and also
astrophysical questions because of the closure of the Z=50 proton shell. The
known isotopes \cite{nndc} cover the range from the proton dripline at $^{100}$Sn
\cite{chartier} to the doubly magic $^{132}$Sn nucleus which is already
$\beta$--unstable. Here, we are mainly interested to investigate the isovector
properties of the D3Y interaction and, as a more general aspect, to test an
effective interaction, determined from symmetric nuclear matter and stable
nuclei, in regions far off stability. No attempt is made to optimize the
parameters in order to obtain a perfect fit of data. The parameter set with
s$_0$=5 is used.

\begin{figure}
\begin{center}
\epsfig{file=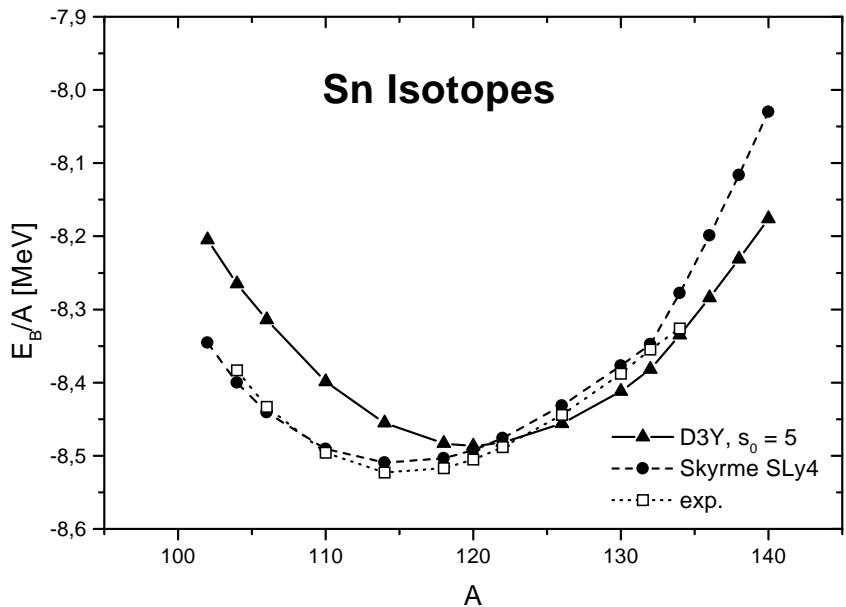, width=14cm}
\caption{
Binding energy per nucleon for the tin isotopes from density dependent
Hartree-Fock calculations using the D3Y interaction with $s_{0} = 5$ (triangles)
and the Skyrme SLy4 force \protect\cite{lyon} (cirles). Experimental binding
energies are denoted by open squares (taken from Ref.
\protect\cite{nndc}). The lines are drawn to guide the eye.}
\label{sn_ener}
\end{center}
\end{figure}

Theoretical and experimental binding energies per particle are compared in Fig.
\ref{sn_ener}. The D3Y binding energies are seen to be shifted slightly to the
left of the data \cite{nndc} but a fair description of the overall dependence on
mass and asymmetry is obtained. The strongest binding is obtained for $^{120}$Sn
while experimentally the minimum is found at approximately $^{115}$Sn. The
binding energies of the neutron--rich isotopes are reasonably well reproduced but
in view of the deviations on the neutron--poor side the agreement might be
fortuitous. Towards the proton dripline, i.e. N$\rightarrow$Z, the binding
energies are increasingly underestimated by up to about 0.12~MeV per nucleon
close to $^{100}$Sn. This result is to some extent unexpected because stable
nuclei with N$\sim$Z were well described. However, also the binding energy of
$^{90}$Zr (Fig. \ref{magic} is underestimated the most and apparently the effect
is enhanced in going from the N=50 to the Z=50 shell. The deviations are in a
range typical for calculations with a G--matrix in finite nuclei. Fully
phenomenological interactions like the Skyrme SLy4 force, for which results are
also displayed in Fig. \ref{sn_ener}, and the Gogny D1 interaction \cite{gogny}
reproduce the Sn binding energies almost perfectly.

Variations of the scaling parameter s$_0$ resulted in an overall shift of the
binding energy curve as it might be expected. Improvements on the neutron--poor
side are obtained on the expense of a worse description of the neutron--rich
isotopes. In particular, the minimum was found to be rather insensitive on
variations of s$_0$. For s$_0$=1 the whole curve is shifted up by about 0.5~MeV
per nucleon with essentially the same shape. These observations lead to the
conclusion that the parameters a$^0_n$ themselves should be readjusted on finite
nuclei with constraints to nuclear matter rather than retaining the nuclear
matter values. Taking the point of view that the nuclear matter parameterization
accounts for the static part of the interaction, where cluster contributions are
included effectively by the fitting procedure \cite{bethe}, such a refinement
could be related to non--static contributions to the in--medium interaction
because of the larger polarizability of the nuclear surface. Theoretically, this
would amount to go beyond the ladder approximation and to include also ring
diagrams. Such contributions are obviously contained in e.g. the Skyrme
parameters but missing in the D3Y interaction. Over the range of the Sn isotopes
the polarization contributions can be expected to decrease towards A$\sim$132
which is known to be a rather stiff nucleus.

\begin{figure}
\begin{center}
\epsfig{file=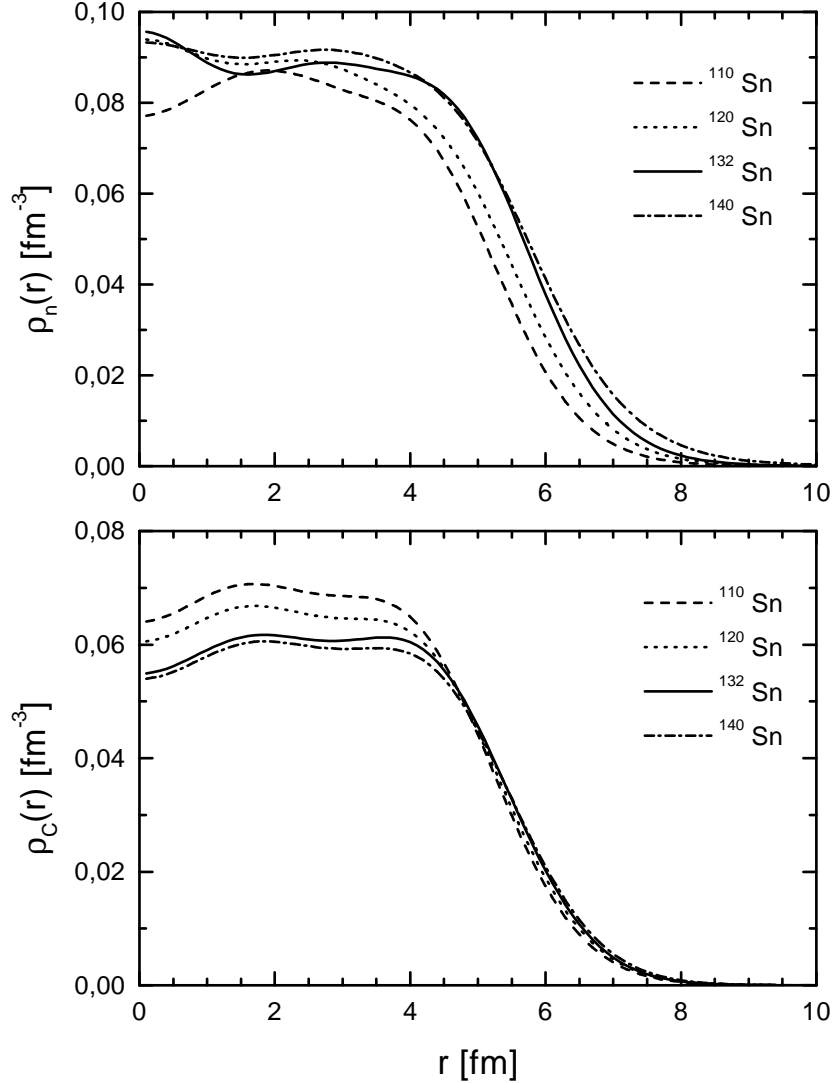, width=12cm}
\caption{
Charge density distributions (lower part) and neutron density distributions
(upper part) of the tin isotopes from density dependent Hartree-Fock calculations
using the D3Y interaction. Results for $s_{0} = 5$ are shown.}
\label{sn_dens}
\end{center}
\end{figure}

\begin{figure}
\begin{center}
\epsfig{file=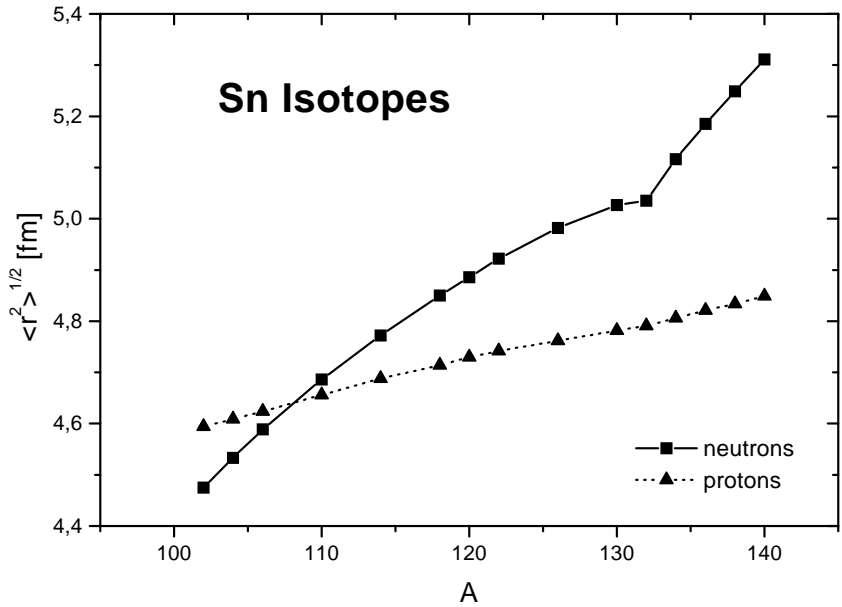, width=14cm}
\caption{
Root mean square radii of the tin isotopes from density dependent Hartree-Fock
calculations using the D3Y interaction with $s_{0} = 5$ for neutrons (squares)
and protons (triangles). The lines are drawn to guide the eye.}
\label{sn_radii}
\end{center}
\end{figure}

\begin{figure}
\begin{center}
\epsfig{file=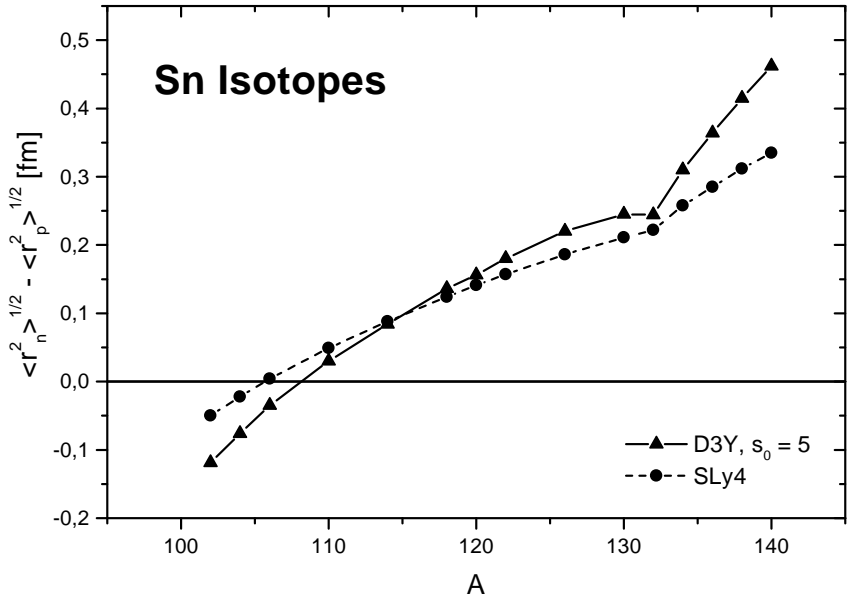, width=14cm}
\caption{
Difference of the root mean square radii of neutrons and protons of the tin
isotopes for the D3Y interaction with $s_{0} = 5$ (triangles) and the Skyrme SLy4
(cirles). The lines are drawn to guide the eye. }
\label{sn_rdif}
\end{center}
\end{figure}

Neutron density distributions and charge densities for several Sn isotopes are
displayed in Fig. \ref{sn_dens}. From $^{110}$Sn to $^{140}$Sn the charge
distributions are slightly reduced in the interior. As seen in Fig.
\ref{sn_radii} this is accompanied by a mild increase of the charge radius by
about 5\%. A more drastic evolution is found for the neutron densities. Beyond
$^{132}$Sn an extremely thick neutron skin is building up leading to a sudden
jump in the neutron rms radii. The neutron skin thickness is more clearly visible
in Fig. \ref{sn_rdif} where the difference of the proton and neutron rms radii is
shown. In $^{140}$Sn a solid layer of neutron matter is found extending by about
0.48~fm beyond the proton matter distribution which is more than a factor of two
larger than the neutron skin in $^{208}$Pb. The same phenomenon is predicted also
by the SLy4 interaction except for a somewhat smaller neutron excess radius. The
saturation of the rms values around A=132 is an indication of the double magic
nature of $^{132}$Sn. The increase is directly related to the shell structure in
the heavy tin isotopes. At A=132 the 2f$_{7/2}$ shell is filled and pairing does
not contribute. At larger masses the neutron 3p--subshells become populated. Weak
binding and the low angular momentum barrier allow a large extension of the
valence wave functions into the exterior. However, the calculations do not lead
to a "halo".

An interesting observation is made from Fig. \ref{sn_rdif} on the neutron--poor
side. Both the D3Y and the SLy4 interactions predict a proton skin for
A$\leq$106. The origins of the proton and neutron skins are quite different.
Atthe neutron rich side the excess neutrons become less bound because of the
increasing repulsion from the isovector potential. Since states with $\ell\geq$1
are involved they are stored predominantly in the surface region of the nucleus.
In the heavy isotopes the isovector potential acts attractively in the proton
sector and thus balances to a large extent the Coulomb repulsion. Since for
N$\rightarrow$Z the isovector part of the mean--field becomes strongly suppressed
the repulsion is no longer compensated and Coulomb effects become visible. Hence,
the proton skin in the neutron--poor Sn nuclides is caused solely by the Coulomb
interaction.

The calculations were extended further into the $\beta$--unstable
region. Up to A=160 the neutron dripline is still not reached. The
separation energies are small but decrease rather slowly.

%-------------------------------------------------------------------------------------
\section{Summary, Discussion and Conclusions} \label{5}

The density matrix expansion (DME) method was used to derive a local energy
density functional for non--relativistic meson exchange interactions. The
interaction energy density was separated into direct and exchange contributions.
The Hartree contributions are treated exactly.	From the systematic expansion of
the one--body density matrix provided by the DME an average momentum density was
determined such that the next to leading order terms in the DME series are
canceled. In uniform systems like infinite nuclear matter all higher order terms
are canceled and the exact energy density is recovered. In finite nuclei the
approach corresponds to a generalized Slater expansion of the density matrix.
Hartree--Fock equations were derived. With our choice for q$^2_F$ a
self--consistently generated effective mass was obtained which is completely
determined by the intrinsic momentum structure of the underlying interaction.

A semi--microscopic approach to the in--medium NN interaction was
discussed. The M3Y G--matrix from the Paris NN potential was modified by
introducing density dependent vertex renormalizations in the isoscalar
and isovector particle--hole channels. The functional form of the vertex
functions was chosen in close analogy to the dynamical structure and the
medium dependence of the vertices in a Brueckner G--matrix. The
parameters, however, were determined empirically by a fit to the binding
energy, equilibrium density, compressibility and the symmetry energy of
symmetric nuclear matter. The momentum structure of the original
G--matrix was retained.

The probably first attempt to parameterize a G--matrix in terms of fixed
propagators and density dependent vertices goes back to Sprung and Banerjee
\cite{SB}. They approximated the momentum structure of the singlet/triplet
even-odd interactions obtained from the Reid NN potential by a superposition of 5
arbitrarily chosen gaussians with channel dependent in--medium vertices. The
functional form of the vertices was chosen in the same way as in section \ref{b}
but using always a first order polynomial. The nuclear matter binding energy
curve of the full Reid G--matrix was reproduced the best for a linear dependence
of the vertices on k$_F$, corresponding to $\beta$=1/3 in Eq.(\ref{vertex_fit}).
From the description of individual matrix elements a slight preference for a
lower value $\beta$=1/6 was deduced \cite{SB}. However, nuclear matter was found
to be badly underbound by about 5~MeV since only two-body correlations were taken
into account. The renormalized vertices obtained in our semi--microscopic
approach include effectively also contributions beyond the two--body cluster
approximation, as was pointed out some time ago by Bethe \cite{bethe}.

In nuclear matter the D3Y interaction leads to an equation of state with
an unusual enhancement of attraction at low densities which is not found
for phenomenological forces. Unfortunately, G--matrix calculations are
not available at such low densities but obviously the free NN T--matrix
should be approached asymptotically. Phenomenological interactions and
also the unrenormalized M3Y G--matrix do not account for this
transition. Without putting too much emphasis on the details of this
particular result it might be expected that such a region of enhanced
attraction could indeed exist in the far tails of nuclear wave functions.
Before drawing final conclusions on this subject a refined treatment of
the low density region is clearly necessary. For that purpose an
extended version of the schematic model for the in--medium vertices
might prove to be useful.

The ground state properties of stable nuclei were reproduced reasonably well. The
good description of binding energies gives confidence to the DME approach,
especially in view of the fact that an interaction from nuclear matter was used.

\vspace{0.5cm}
{\bf Acknowledgement:}

This work was supported in part by DFG (contract Le439/4-1), GSI and BMBF.

%-------------------------------------------------------------------------------------
\begin{appendix}
%-------------------------------------------------------------------------------------
\section{Representations Of The NN Interaction} \label{app}

In literature several representations of the NN interaction exist which can be
found in, e. g. \cite{bertsch}. Here we will only consider the central part of
the interaction. The M3Y interaction is given in a representation specifying the
two-body channel spin and the parity (SE,TE,SO,TO) which is commonly used in
inelastic scattering. In our calculations we are especially interested in the p-p
resp. p-n interaction for calculating the Hartree-Fock matrix elements, the (pp,
pn) representation. It can be easily obtained with the transformation
\cite{negalt}
\begin{eqnarray}
	V_{pp}^{d}	    & = & \frac{1}{8} \left[ 2 V_{SE} + 6 V_{TO} \right]
	= V_{nn}^{d} \nonumber \\ V_{pn}^{d} & = & \frac{1}{8} \left[ 1 V_{SE} +
	3 V_{TE} + 1 V_{SO} + 3 V_{TO} \right] = V_{np}^{d} \nonumber \\
	V_{pp}^{e} & = & \frac{1}{8} \left[ 2 V_{SE} - 6 V_{TO} \right] =
	V_{nn}^{e} \nonumber \\ V_{pn}^{e}	    & = & \frac{1}{8} \left[ 1
	V_{SE} + 3 V_{TE} - 1 V_{SO} - 3 V_{TO} \right] = V_{np}^{e}
\end{eqnarray}
Here, the indices $d$ resp. $e$ stand for the direct resp. exchange part of the
NN interaction. A representation especially useful if one uses a contact
interaction or is interested in the direct part of the interaction is the set
($V_{0},V_{\sigma},V_{\tau},V_{\sigma \tau}$), which follows from the set
(SE,TE,SO,TO) through
\begin{eqnarray}
	V_{0}		& = & \frac{1}{16} \left[ 3 V_{SE} + 3 V_{TE} + 1 V_{SO}
	+ 9 V_{TO} \right] \nonumber \\ V_{\sigma}	& = & \frac{1}{16} \left[
	-3 V_{SE} + 1 V_{TE} - 1 V_{SO} + 3 V_{TO} \right] \nonumber \\ V_{\tau}
	& = & \frac{1}{16} \left[ 1 V_{SE} - 3 V_{TE} - 1  V_{SO} + 3 V_{TO}
	\right] \nonumber \\ V_{\sigma \tau} & = & \frac{1}{16} \left[ -1 V_{SE}
	- 1 V_{TE} + 1 V_{SO} + 1 V_{TO} \right]
\end{eqnarray}
Our aim is to find a spin averaged interaction which separates the isoscalar and
the isovector interaction for both the direct and the exchange part. Defining
\begin{eqnarray} \nonumber
    V_{0}^{d/e} & = & \frac{1}{4} (V_{pp}^{d/e} + V_{pn}^{d/e} + V_{np}^{d/e} +
    V_{nn}^{d/e}) \\ V_{\tau}^{d/e} & = & \frac{1}{4} (V_{pp}^{d/e} -
    V_{pn}^{d/e} - V_{np}^{d/e} + V_{nn}^{d/e}) \label{eapp1}
\end{eqnarray}
we find in the (SE,TE,SO,TO) representation
\begin{eqnarray}
	V_{0}^{d}	    & = & \frac{1}{16} \left[ 3 V_{SE} + 3 V_{TE} + 1
	V_{SO} + 9 V_{TO} \right] = V_{0} \nonumber \\ V_{\tau}^{d}	 & = &
	\frac{1}{16} \left[ 1 V_{SE} - 3 V_{TE} - 1 V_{SO} + 3 V_{TO} \right] =
	V_{\tau} \nonumber \\ V_{0}^{e} & = & \frac{1}{16} \left[ 3 V_{SE} + 3
	V_{TE} - 1  V_{SO} - 9 V_{TO} \right] \nonumber \\ V_{\tau}^{e} & = &
	\frac{1}{16} \left[ 1 V_{SE} - 3 V_{TE} + 1 V_{SO} - 3 V_{TO} \right] .
\end{eqnarray}
One sees that the defined direct terms of this representation are equivalent to
the ones of the set ($V_{0},V_{\sigma},V_{\tau},V_{\sigma \tau}$) but also that
there is no equivalent for the exchange terms.

\end{appendix}

%-------------------------------------------------------------------------------------

%-------------------------------------------------------------------------------------
\end{document}